\documentclass[11pt]{article}

\usepackage[margin=2cm]{geometry}
\usepackage{booktabs,tabularx,float}
\usepackage{tcolorbox}
\usepackage[hidelinks,colorlinks=true,linkcolor=blue!60!black,urlcolor=blue!60!black,citecolor=blue!60!black]{hyperref}
\usepackage{enumitem}
\usepackage{xcolor}
\usepackage{colortbl}
\usepackage{array}
\usepackage{times}
\usepackage{amsmath,amssymb}
\usepackage{graphicx}
\usepackage{caption}
\usepackage{subcaption}
\usepackage{microtype}
\usepackage{pifont}
\usepackage{url}

\tcbuselibrary{skins,breakable}

\graphicspath{{./}}

\newcolumntype{L}[1]{>{\raggedright\arraybackslash}p{#1}}
\newcolumntype{R}[1]{>{\raggedleft\arraybackslash}p{#1}}
\newcolumntype{C}[1]{>{\centering\arraybackslash}p{#1}}

\definecolor{lightgray}{RGB}{245,245,245}
\definecolor{promptblue}{RGB}{21,101,192}
\definecolor{promptteal}{RGB}{0,77,64}
\definecolor{passfill}{RGB}{232,245,233}
\definecolor{failfill}{RGB}{255,235,238}

\newcommand{\cmark}{\ding{51}}
\newcommand{\xmark}{\ding{55}}

\newtcolorbox{promptbox}[1]{
  colback=white, colframe=promptblue, boxrule=0.8pt, arc=3pt,
  fonttitle=\bfseries, title={#1}, breakable
}
\newtcolorbox{amrbox}[1]{
  colback=white, colframe=promptteal, boxrule=0.8pt, arc=3pt,
  fonttitle=\bfseries, title={#1}, breakable
}

\title{\vspace{-1.5cm}\textbf{SWE-WebDevBench: Evaluating Coding Agent\\ Application Platforms as Virtual Software Agencies}}

\author{
Siddhant Saxena$^{1}$ \quad
Nilesh Trivedi$^{2}$ \quad
Vinayaka Jyothi$^{2}$ \\[0.3em]
{\small $^{1}$BaseThesis Labs \quad $^{2}$QwikBuild}
}

\date{\vspace{-0.5em}March 2026}

\begin{document}
\maketitle

\begin{tcolorbox}[colback=lightgray,colframe=black!30,boxrule=0.4pt,arc=2pt,left=6pt,right=6pt,top=4pt,bottom=4pt]
\footnotesize
\begin{center}\textbf{Abstract}\end{center}
\vspace{-0.7em}

The emergence of ``vibe coding''~\cite{karpathy2025vibe} platforms, where users describe applications in natural language and AI agents autonomously generate full-stack software, has created a need for rigorous evaluation beyond code-level benchmarks. Existing frameworks such as SWE-bench and FeatBench evaluate coding agents on developer-centric tasks but do not assess whether an AI platform can function as a \emph{complete software agency}: understanding business requirements, making architectural decisions, writing production code, handling iterative modifications, and maintaining business readiness.
\\

We introduce \textbf{SWE-WebDev Bench}, a 68-metric evaluation framework spanning 25 primary and 43 diagnostic metrics across seven groups, organized along three dimensions: Interaction Mode (App Creation Request (ACR) vs.\ App Modification Request (AMR)), Agency Angle (Product Manager (PM), Engineering, Ops), and Complexity Tier (T4 multi-role SaaS, T5 AI-native). We evaluate six platforms using six standardized prompts across three business domains with 80 embedded canary requirements.
\\

Our initial evaluation (six platforms, three domains, 18 evaluation cells) reveals four recurring shortcomings in the current generation of AI app builders: (1)~A \emph{specification bottleneck}, where platforms compress rich business requirements into oversimplified technical plans, with observed inference quality varying widely across platforms (Canary Retention Rate: 17.7\%--97.7\%). (2)~A pervasive \emph{frontend-backend decoupling}, where visually polished UIs mask absent or broken backend infrastructure, with Background Job Scores ranging from 0\% to 49\%. (3)~A steep \emph{production readiness cliff}, where no platform scores above 60\% on engineering quality and post-generation human effort varies substantially across platforms. (4)~\emph{Widespread security and infrastructure failures}, with no platform exceeding 65\% Security Score against a 90\% target and concurrency handling as low as 6\%. These observations are descriptive of our sample and require larger-scale replication to establish generality. We release SWE-WebDev Bench as a community benchmark to enable such replication and help platform builders identify and address these gaps. 
\\

\textbf{Code and benchmark resources are available at:} \url{https://github.com/snowmountainAi/webdevbench} and \url{https://webdevbench.com/}.
\\

\noindent\textit{Disclosure: Two authors are affiliated with QwikBuild, one of the six evaluated platforms. This creates a potential conflict of interest. To mitigate this, we release all prompts, rubrics, and scoring protocols for independent replication and describe bias mitigation measures in \S\ref{sec:limitations}. We encourage independent evaluations by parties unaffiliated with any evaluated platform.}
\end{tcolorbox}

\vspace{2pt}

\begin{figure}[h!]
\centering
\includegraphics[width=0.95\textwidth]{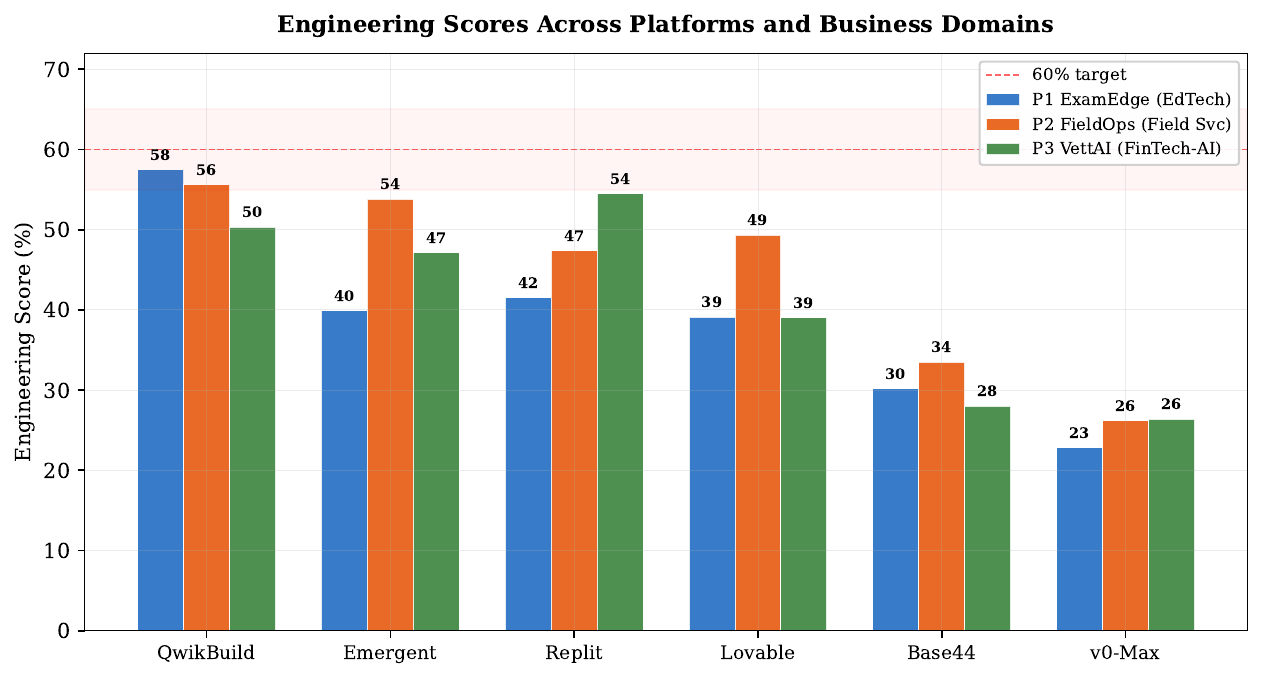}
\vspace{-0.5em}
\caption{\small Engineering Scores on SWE-WebDev Bench across six platforms and three business domains. No platform exceeds 60\%, indicating substantial room for improvement across the field. Domain-specific performance swings (e.g., 13-point variance for Replit between P1 and P3) suggest that current platforms lack generalized competence.}
\label{fig:main_result}
\end{figure}

\section{Introduction}
\label{sec:intro}

``The hottest new programming language is English''~\cite{karpathy2023}. This observation has materialized into a market of AI application-building platforms where users describe software in natural language and receive deployed, full-stack applications. Platforms such as Lovable, Replit Agent, Vercel v0, and others claim to compress months of development into minutes, making software creation accessible to non-developers.

Yet the question of \emph{quality}---whether the generated software is actually production-ready---remains largely unanswered by the research community. Existing evaluation frameworks fall into three categories, none of which addresses the full scope of what these platforms claim to deliver.

\textbf{Code-level benchmarks} such as HumanEval~\cite{humaneval} and ClassEval~\cite{classeval} evaluate function-level code generation and require code-level specifications as input, which is incompatible with the vibe coding paradigm where users provide only natural language.

\textbf{Issue-solving benchmarks} such as SWE-bench~\cite{swebench} and its successors evaluate the ability to produce patches from issue descriptions. FeatBench~\cite{featbench} extends this to feature implementation, finding that even the best agent (GPT-5 with Trae-agent~\cite{trae_agent}) resolves only 29.94\% of tasks. However, these benchmarks evaluate developer-facing scenarios on existing codebases and do not assess whether an AI system can build a complete application from scratch for a non-technical user.

\textbf{Emerging application-level evaluations} have begun addressing whole-application generation. Vibe Code Bench~\cite{vibecodebench} evaluates end-to-end web app generation using browser-based workflow testing on 100 specifications, but does not assess PM behavior or iterative modification. WebCoderBench~\cite{webcoderbench} introduces 24 fine-grained metrics across 1,572 real user requirements, but evaluates single-page applications without deployment or security assessment. From Prompt to Product~\cite{prompttoproduct} conducts a human-centered comparison of three commercial platforms (Replit, Bolt, Firebase Studio) using 205 participants, but relies on pairwise preference judgments rather than metric-level diagnostics. WebGen-Bench~\cite{webgenbench} tests multi-file website generation across 647 test cases, finding that even the best agent achieves only 27.8\% accuracy. These benchmarks represent important progress toward application-level evaluation but do not assess the full agency pipeline: requirement elicitation, iterative modification, security, infrastructure, and business readiness.

The gap these frameworks leave unfilled is the scenario that matters most for the vibe coding paradigm: \emph{Can an AI platform function as a complete software agency, understanding business intent, clarifying ambiguities, making sound architectural decisions, writing secure code, and handling iterative modifications?}

We introduce \textbf{SWE-WebDev Bench}, an evaluation framework designed to answer this question and, more importantly, to surface the specific failure modes that the community must address. Our contributions are:

\begin{enumerate}[leftmargin=*,itemsep=2pt]
\item A \textbf{68-metric evaluation framework} organized across three orthogonal dimensions (Mode $\times$ Angle $\times$ Tier) with a four-tier judging taxonomy, designed to diagnose where AI app builders fall short of production readiness (\S\ref{sec:framework}).

\item The \textbf{ACR/AMR distinction}: the first benchmark to separately evaluate App Creation Requests and App Modification Requests, revealing that modification handling is a fundamentally different and harder competency (\S\ref{sec:design}).

\item The \textbf{Canary Requirement methodology}: 80 culturally-specific, domain-embedded test requirements with four types (Original, New, Surviving, Contradiction) that distinguish genuine comprehension from template matching (\S\ref{sec:canary}).

\item An \textbf{initial six-platform evaluation} across three business domains, revealing four recurring shortcomings observed across all evaluated platforms (\S\ref{sec:results}).

\item We release SWE-WebDev Bench, including prompts, rubrics, and evaluation protocols, to enable independent replication and benchmarking (\url{https://github.com/snowmountainAi/webdevbench}, \url{https://webdevbench.com/}).
\end{enumerate}

\section{Related Work}
\label{sec:related}

Table~\ref{tab:benchmark_comparison} positions SWE-WebDev Bench against existing evaluation approaches.

\begin{table}[htbp]
\centering
\caption{Comparison with existing evaluation frameworks for AI coding systems.}
\label{tab:benchmark_comparison}
\small
\begin{tabular}{@{}L{2.8cm}L{2.2cm}C{1.2cm}C{1.4cm}C{1.2cm}C{1.5cm}@{}}
\toprule
\textbf{Benchmark} & \textbf{Scope} & \textbf{Eval PM?} & \textbf{Code Eval} & \textbf{AMR?} & \textbf{Platforms} \\
\midrule
HumanEval~\cite{humaneval} & Function-level & \xmark & Partial & \xmark & Models \\
SWE-bench~\cite{swebench} & Issue patches & \xmark & Patch only & \xmark & Models \\
FeatBench~\cite{featbench} & Feature impl. & \xmark & Patch only & \xmark & Agents \\
Vibe Code Bench~\cite{vibecodebench} & Web apps & \xmark & Browser & \xmark & Models \\
WebCoderBench~\cite{webcoderbench} & Web apps & \xmark & 24 metrics & \xmark & Models \\
Prompt$\to$Product~\cite{prompttoproduct} & App platforms & \xmark & Human pref. & \xmark & 3 \\
WebGen-Bench~\cite{webgenbench} & Websites & \xmark & Functional & \xmark & Agents \\
\midrule
\textbf{SWE-WebDev Bench} & \textbf{Full agency} & \textbf{\cmark\,(68m)} & \textbf{Full audit} & \textbf{\cmark} & \textbf{6} \\
\bottomrule
\end{tabular}
\end{table}

\subsection{Code Generation Benchmarks}

The trajectory from HumanEval~\cite{humaneval} through MBPP~\cite{mbpp} to ClassEval~\cite{classeval} represents a progressive broadening of code generation evaluation from function-level to class-level tasks. However, all these benchmarks require code-level specifications (function signatures, docstrings) as input, which is incompatible with vibe coding where the user provides only natural language intent.

SWE-bench~\cite{swebench} moved closer to realistic scenarios by tasking agents with resolving real GitHub issues. SWE-bench-Live~\cite{swebenchlive} added temporal freshness, and FeatBench~\cite{featbench} shifted focus from issue-solving to feature implementation. FeatBench's key finding, that 73.6\% of failures stem from regressive implementation where the agent breaks existing functionality while adding features, directly motivates our AMR evaluation dimension.

A common limitation across all these benchmarks is that they evaluate patch quality on existing codebases, not complete application delivery. They assume a developer audience, test single-file or few-file changes, and do not address the PM, deployment, or business-readiness dimensions that define real software delivery.

\subsection{Web Application Generation Benchmarks}

A recent wave of benchmarks has begun evaluating AI systems on whole-application generation. Vibe Code Bench~\cite{vibecodebench} evaluates 100 web application specifications using 964 browser-based workflows with 10,131 substeps, finding that the best model achieves 61.8\% accuracy---a much more discriminative benchmark than SWE-bench (42.7\% vs.\ 2.8\% gap between top and bottom models). WebCoderBench~\cite{webcoderbench} introduces 24 fine-grained evaluation metrics across 9 perspectives for 1,572 real user requirements, incorporating user-preference-weighted scoring. WebGen-Bench~\cite{webgenbench} evaluates multi-file website generation from scratch across 647 test cases, where even the best agent (Bolt.diy + DeepSeek-R1) achieves only 27.8\% accuracy. From Prompt to Product~\cite{prompttoproduct} takes a human-centered approach, evaluating three commercial platforms (Replit, Bolt, Firebase Studio) using 96 prompts and 205 human participants with 1,071 pairwise comparisons. FullStack Bench~\cite{fullstackbench} evaluates full-stack coding across 16 languages and 11 domains with 3,374 problems, but tests isolated problems rather than coherent applications. SUSVIBES~\cite{susvibes} specifically benchmarks the security of agent-generated code, finding that while 61\% of solutions are functionally correct, only 10.5\% are secure---directly supporting our Finding~4 on universal security failures.

These benchmarks represent important progress but share common gaps that SWE-WebDev Bench addresses: none evaluates requirement elicitation (PM behavior), none measures iterative modification handling (AMR), and none assesses the full pipeline from business intent through deployment readiness. Our framework is complementary: where Vibe Code Bench measures \emph{whether} the app works, SWE-WebDev Bench measures \emph{why} it fails and \emph{what} to fix.

\subsection{Visual and Multimodal Evaluation}

We observe an analogous phenomenon in AI app builders: a \emph{specification bottleneck} where platforms compress rich, ambiguous business requirements into oversimplified technical plans, losing critical domain context. Our PM Agent evaluation dimension (\S\ref{sec:pm_results}) directly measures this compression loss.

\subsection{Agent Evaluation Methodology}

Cognition AI's blog post on evaluating coding agents~\cite{cognition_eval} introduced realistic environments with simulated users and evaluator agents for autonomous outcome assessment. Their concept of ``interactive self-reflection,'' where agents use environment signals to evaluate themselves, informs our Ops/Maintenance evaluation angle. However, Cognition evaluates a single agent on developer tasks. Our framework evaluates six platforms on business-user tasks across a multi-dimensional quality space.

\subsection{LLM-as-Judge Evaluation}

SWE-WebDev Bench relies on LLM judges for Tier~1 and Tier~2 metrics, which places it within the growing literature on automated evaluation. Zheng et al.~\cite{zheng2023judging} introduced MT-Bench and demonstrated that strong LLMs can approximate human preferences with $>80\%$ agreement, but also identified systematic biases: position bias (favoring the first option), verbosity bias (favoring longer outputs), and self-enhancement bias (favoring outputs from the same model family). Kim et al.~\cite{kim2024prometheus} showed that fine-tuned judge models can achieve higher correlation with human evaluators when given detailed rubrics, motivating our structured scoring rubrics for each metric. Li et al.~\cite{li2024generativejudge} further documented that LLM judges exhibit platform-specific biases when evaluating code, an important consideration given that our evaluation targets commercial platforms with distinctive code styles.

We address these concerns through our tiered approach: high-stakes subjective metrics (BIF, ETF, FGD) are assigned to Tier~3 expert panels rather than LLM judges, while LLM judges are used for factual verification tasks (Tier~1: ``does this API route exist?'') where bias is minimal. We report measured inter-rater agreement in \S\ref{sec:framework} and discuss calibration limitations in \S\ref{sec:limitations}.

\subsection{Benchmark Governance and Maintenance}

The challenge of benchmark maintenance and community governance has been addressed by several large-scale evaluation efforts. HELM~\cite{liang2023helm} established a model for living benchmarks with regular re-evaluation, transparent methodology, and community contribution protocols. Dynabench~\cite{kiela2021dynabench} introduced dynamic, adversarial benchmarking to resist saturation. Chatbot Arena~\cite{zheng2023judging} demonstrated that community-driven pairwise evaluation can scale to thousands of comparisons. SWE-WebDev Bench draws on these precedents in its governance plan (\S\ref{sec:governance}).

\section{The SWE-WebDev Bench Framework}
\label{sec:framework}

\subsection{Design Principles}

The design of SWE-WebDev Bench is guided by four principles that address specific limitations we identified in existing evaluation approaches.

\textbf{Principle 1: Evaluate the full delivery pipeline, not just the code.} When a non-technical user asks an AI platform to build a SaaS application, the platform must perform the work of an entire software agency: a product manager who interprets ambiguous requirements, engineers who write correct and secure code, and an operations team who deploys and maintains the result. Existing benchmarks evaluate only the engineering phase (code patches, function implementations). SWE-WebDev Bench evaluates all three phases, because failure in any one of them renders the output unusable for the target user.

\textbf{Principle 2: Measure what the user cannot verify.} The vibe coding paradigm shifts software creation to users who cannot read code. This creates a unique evaluation challenge: the most dangerous failures are \emph{invisible} ones---silent specification violations (a date format quietly defaulting to MM/DD/YYYY instead of the requested DD/MM/YYYY), security vulnerabilities in generated backend code, or regression bugs introduced during modification. SWE-WebDev Bench prioritizes metrics that surface these invisible failures, because they represent the gap between perceived and actual quality.

\textbf{Principle 3: Diagnose, not just score.} A benchmark that reports ``Platform X scored 47\%'' is useful for ranking but not for improvement. SWE-WebDev Bench pairs every primary metric (which measures \emph{what} was delivered) with diagnostic metrics (which trace \emph{why} it succeeded or failed). This dual structure is designed to make the benchmark actionable for platform builders: a low Feature Completeness Score can be traced to poor requirement capture, hallucinated features, or implementation failures---each demanding a different architectural intervention.

\textbf{Principle 4: Resist gaming through specificity.} AI benchmarks are vulnerable to overfitting: platforms can optimize for benchmark-specific patterns without improving general capability. We resist this through three mechanisms: (a) canary requirements that are culturally embedded and domain-specific, making them difficult to hard-code; (b) deliberately varied prompt styles (stream-of-consciousness, formal RFP, technical specification) that prevent optimization for a single input format; and (c) the ACR/AMR distinction, which requires genuine code understanding rather than template-based generation.

\subsection{Evaluation Cube: Three Orthogonal Dimensions}

\begin{figure}[htbp]
\centering
\includegraphics[width=0.95\textwidth]{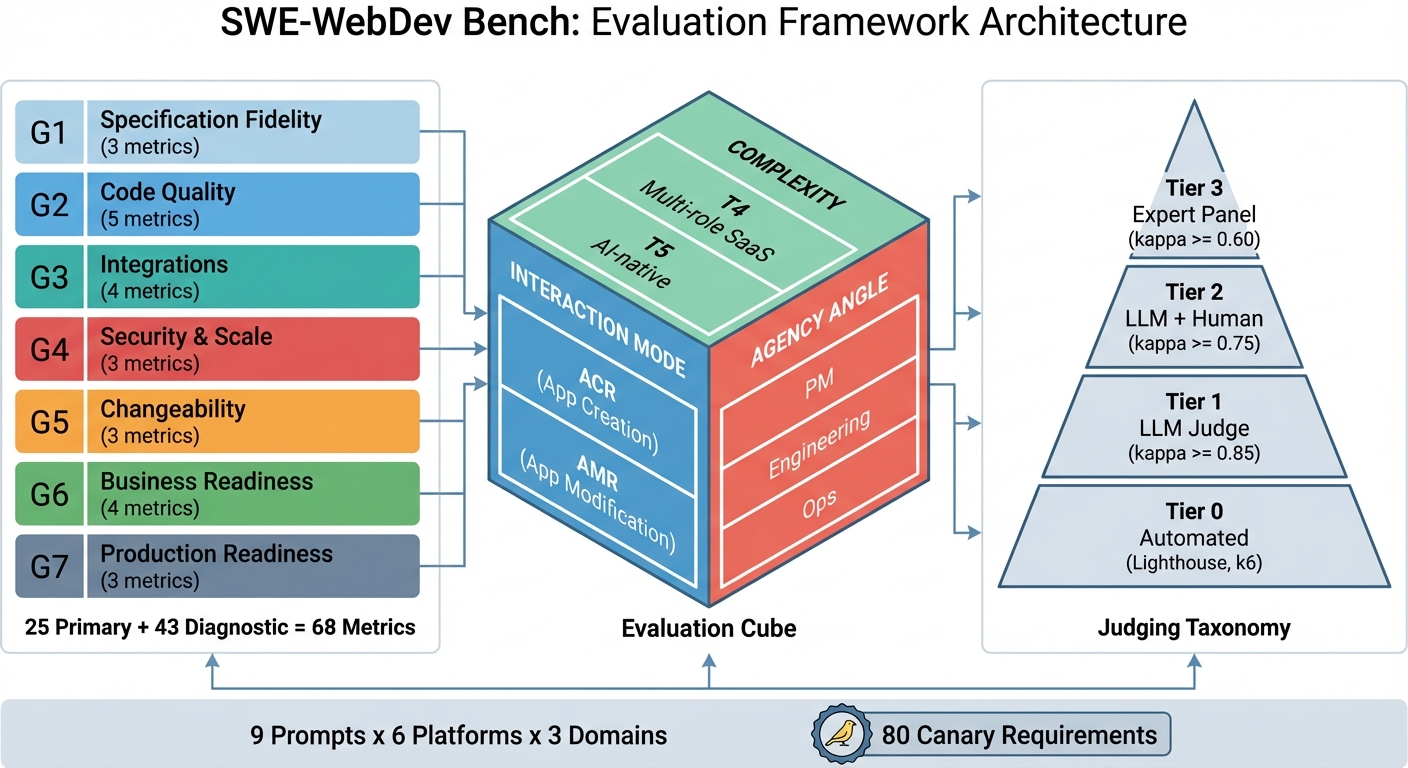}
\vspace{-0.5em}
\caption{\small SWE-WebDev Bench evaluation framework architecture. \textbf{Left:} Seven metric groups (G1--G7) spanning 25 primary and 43 diagnostic metrics. \textbf{Center:} The Evaluation Cube with three orthogonal dimensions---Interaction Mode (ACR/AMR), Agency Angle (PM/Engineering/Ops), and Complexity Tier (T4/T5). \textbf{Right:} Four-tier judging taxonomy from fully automated (Tier~0) to expert panel (Tier~3). \textbf{Bottom:} 80 canary requirements embedded across 9 prompts, 6 platforms, and 3 domains.}
\label{fig:framework}
\end{figure}

These principles are operationalized through three orthogonal dimensions that form an evaluation cube (Figure~\ref{fig:framework}). Each dimension was chosen to isolate a specific axis of variation that existing benchmarks collapse.

\textbf{Dimension 1: Interaction Mode (ACR vs.\ AMR).} We distinguish \textit{App Creation Requests} (ACR), where the platform builds a new application from natural language, and \textit{App Modification Requests} (AMR), where the platform must modify an existing application while preserving functionality.

\textit{Why this dimension matters:} Vibe coding is inherently iterative---users rarely describe their complete application in a single prompt. They build, use, and then request changes (``add multi-tenancy,'' ``swap the AI provider,'' ``the dispatch system needs to be smarter''). AMR is strictly harder than ACR because it requires understanding existing code, managing regressions, and scoping changes precisely. FeatBench~\cite{featbench} found that 73.6\% of coding agent failures on modification tasks involve breaking existing functionality, confirming that creation and modification are fundamentally different competencies that must be evaluated separately. No existing platform benchmark distinguishes these modes.

\textbf{Dimension 2: Agency Angle (PM $\times$ Engineering $\times$ Ops).} We decompose platform quality into three roles that mirror a human software agency: \textit{Product Manager (PM)} for requirement understanding, inference, ambiguity handling, and plan quality; \textit{Engineering (E)} for code quality, architecture, integrations, security, and AI feature implementation; and \textit{Operations (O)} for deployment, monitoring, stability, and performance.

\textit{Why this dimension matters:} When a human software agency delivers a project, failures can be traced to a specific role: the PM misunderstood the client, the engineers wrote buggy code, or operations failed to deploy reliably. AI app-building platforms bundle all three roles into a single system, making it difficult to diagnose where quality breaks down. By tagging each metric with its applicable agency angle, SWE-WebDev Bench enables targeted diagnosis. Our results validate this decomposition: the PM angle shows the widest variance across platforms (3.5$\times$ on Inference Quality Score), while Engineering scores are more compressed (6-point spread on Frontend Engineering), suggesting that PM capability---not code generation---is the primary differentiator.

\textbf{Dimension 3: Complexity Tier (T4 vs.\ T5).} We evaluate at two tiers: T4 (multi-role Software-as-a-Service (SaaS) with Role-Based Access Control (RBAC), scheduled jobs, multiple integrations) and T5 (AI-native multi-tenant applications with LLM pipelines, trust/safety constraints, and provider abstraction).

\textit{Why this dimension matters:} Complexity tiers prevent a common evaluation pitfall: a platform that builds excellent to-do apps may fail entirely on applications requiring role-based access control, background job scheduling, or AI pipeline orchestration. T4 represents the complexity floor for real business applications (most SaaS products require RBAC and integrations). T5 adds the emerging dimension of AI-native applications where the generated code must itself orchestrate LLM calls safely and reliably. By evaluating at both tiers, SWE-WebDev Bench reveals whether platforms scale gracefully or hit capability cliffs as application complexity increases.

\subsection{Metric Taxonomy: 25 Primary + 43 Diagnostic}

SWE-WebDev Bench comprises 68 metrics: 25 primary metrics across 7 groups (Table~\ref{tab:primary_metrics}) and 43 diagnostic metrics across 4 categories.

\begin{table}[htbp]
\centering
\caption{Primary Metric Groups (25 metrics). Each metric is tagged with applicable Agency Angle(s). Full metric names are provided; abbreviations are used in subsequent tables.}
\label{tab:primary_metrics}
\small
\begin{tabular}{@{}L{0.6cm}L{7.5cm}cL{1.8cm}L{4cm}@{}}
\toprule
\textbf{Grp} & \textbf{Metrics} & \textbf{\#} & \textbf{Angle} & \textbf{Key Signal} \\
\midrule
G1 & Business Intent Fidelity (BIF), Feature Completeness Score (FCS), Canary Retention Rate (CRR) & 3 & PM & Did it understand what was asked? \\
G2 & Schema Design Score (SDS), Backend Logic Score (BLS), Frontend Engineering Score (FES), Code Hygiene Score (CHS), Architecture Score (ARC) & 5 & Engineering & Is the generated code well-built? \\
G3 & Core Integration Score (CIS), AI-Inside-App Score (AIA), External Service Reliability (ESR), Cron \& Background Jobs Score (CBS) & 4 & E + Ops & Do integrations and jobs work? \\
G4 & Security Score (SS), Scalability Architecture Score (SAS), Concurrency \& Load Score (CLS) & 3 & E + Ops & Is it secure and scalable? \\
G5 & Code Change Impact Score (CCIS), Effort-to-Fix (ETF), Post-PRD Human Effort (PHE) & 3 & E + Ops & Can it handle change without breaking? \\
G6 & SEO \& Web Standards Score (SWS), Lead \& Growth Score (LGS), Multilingual/Localisation Score (MLS), AI Feature Quality (AFQ) & 4 & E + PM & Is it business-ready? \\
G7 & Total Cost of Correctness (TCC), Feature Gap Delta (FGD), Claim Drift Index (CDI) & 3 & PM + Ops & What does it actually cost to ship? \\
\bottomrule
\end{tabular}
\end{table}

\subsubsection{Metric Group Rationale}

The seven groups are designed so that each captures a distinct failure mode that existing benchmarks miss. Together, they cover the full lifecycle of software delivery---from understanding what to build, through building it correctly, to shipping and maintaining it in production.

\textbf{G1: Specification Fidelity} measures whether the platform understood what the user asked for. Existing code benchmarks (HumanEval, SWE-bench) take specifications as given; in vibe coding, the specification itself must be inferred from ambiguous natural language. Business Intent Fidelity (BIF) captures whether the platform grasps the user's business purpose, not just their literal words. Feature Completeness Score (FCS) measures functional coverage against a reference specification. Canary Retention Rate (CRR) is methodologically novel: by embedding culturally-specific requirements (e.g., DD/MM/YYYY date format, INR currency, JEE/NEET exam conventions) that are easy for template-matching systems to drop, CRR distinguishes genuine comprehension from shallow pattern extraction. A platform with high FCS but low CRR is building the right features with the wrong details---a failure mode invisible to existing benchmarks.

\textbf{G2: Code Quality} evaluates the engineering quality of generated code across five dimensions. Schema Design Score (SDS) assesses database modeling---normalization, referential integrity, indexing, and multi-tenancy support---because poor schema design is the single most expensive technical debt category in web applications. Backend Logic Score (BLS) evaluates API design, route structure, and business logic correctness. Frontend Engineering Score (FES) measures component architecture, state management, and UI/UX implementation. Code Hygiene Score (CHS) captures maintainability factors: naming conventions, dead code, duplication, and separation of concerns. Architecture Score (ARC) assesses overall system design: separation of layers, dependency management, and pattern consistency. This five-dimensional decomposition is necessary because our results show that platforms can score 70\%+ on FES while scoring below 10\% on SDS (Table~\ref{tab:per_prompt})---a granularity that composite ``code quality'' scores would mask.

\textbf{G3: Integrations} measures whether the platform can connect to real-world services and implement background processing---capabilities that separate prototypes from production applications. Core Integration Score (CIS) tests database CRUD (Create, Read, Update, Delete) operations, authentication flows, and file storage. AI-Inside-App Score (AIA) evaluates AI feature implementation quality: prompt engineering, error handling, provider abstraction, and trust/safety controls. External Service Reliability (ESR) tests third-party service integration (email, SMS, payment). Cron \& Background Jobs Score (CBS) measures scheduled task implementation. We include this group because our results reveal it as the highest-variance capability across platforms: CBS ranges from 0\% to 49\%, a 50-point spread that determines whether an application can run autonomously in production.

\textbf{G4: Security \& Scale} addresses the non-functional requirements that determine production viability. Security Score (SS) covers OWASP (Open Web Application Security Project) Top 10 vulnerabilities, authentication hardening, API key management, and access control. Scalability Architecture Score (SAS) evaluates connection pooling, caching strategy, and horizontal scaling readiness. Concurrency \& Load Score (CLS) uses the k6 load testing tool to measure behavior under concurrent users. We set aggressive targets (SS $\geq$ 90\%, SAS $\geq$ 70\%, CLS $\geq$ 70\%) because production deployment of insecure or non-scalable applications poses real risk to end users. Our results confirm this group as a universal failure point: no platform exceeds 65\% on Security Score.

\textbf{G5: Changeability} is unique to SWE-WebDev Bench and directly motivated by the iterative nature of vibe coding. Users do not build an application once; they iterate: ``add a feature,'' ``change this,'' ``now support Hindi.'' Code Change Impact Score (CCIS) measures whether modifications break existing functionality---the regression problem that FeatBench~\cite{featbench} found affects 73.6\% of coding agent modifications. Effort-to-Fix (ETF) estimates the developer-hours required to bring the generated application to production quality. Post-PRD (Product Requirements Document) Human Effort (PHE) counts the re-prompts and manual code edits needed after initial generation. These metrics quantify the ``last mile'' problem: how much human effort remains after the AI has done its work.

\textbf{G6: Business Readiness} captures whether the application is ready for real users and real markets, a dimension entirely absent from existing code benchmarks. SEO \& Web Standards Score (SWS) tests meta tags, sitemaps, structured data, and Core Web Vitals compliance---requirements that determine whether the application is discoverable. Lead \& Growth Score (LGS) evaluates analytics integration, onboarding flows, and conversion tracking. Multilingual/Localisation Score (MLS) tests internationalization (i18n) support, a critical requirement for applications targeting non-English-speaking markets. AI Feature Quality (AFQ) evaluates the quality of AI-powered features within the application (distinct from AIA, which measures integration correctness). These metrics matter because vibe coding's promise is to deliver business-ready software, not just technically functional code.

\textbf{G7: Production Readiness} provides cost-oriented metrics that directly inform platform selection decisions. Total Cost of Correctness (TCC) aggregates platform costs, compute time, and human effort into a single monetary figure. Feature Gap Delta (FGD) estimates developer-hours needed to implement missing features. Claim Drift Index (CDI) measures the gap between what the platform claims to have built (in its PRD, traces, and dashboard) and what actually works---a trust metric. CDI is particularly important for vibe coding because non-technical users rely on the platform's self-reported status; if CDI is high, the user believes the application is complete when it is not, leading to failed deployments and lost trust.

\subsubsection{Diagnostic Metrics}

The 43 diagnostic metrics provide process-level insight into \emph{why} platforms succeed or fail, complementing the primary metrics that measure \emph{what} was delivered. They are organized into four categories. Category~A (PRD \& Planning) covers intent capture (Explicit Capture Rate, Inference Quality Score, Hallucination Rate, Inference Precision, Critical Omission Rate), conversational guidance (Conversational Guidance Score, Question Efficiency Rate, Conversational Turns to Convergence, Proactive Discovery Rate), ambiguity handling (Calibrated Uncertainty Score, Under-Clarification Rate, Over-Clarification Rate), and plan quality (Plan Structural Validity, Plan Communicability Score). Category~B (Build \& Code) covers trace clarity, error visibility, canary deep-dive, and---critically for AMR---change management diagnostics (Adaptive Coherence Score, Regression Rate, Change Acknowledgment Rate, Plan Update Rate, Change Processing Overhead). Category~C (Deploy \& Maintenance) covers deployment success, time to first byte, uptime, and post-deploy stability. Category~D (Operator Experience) covers claim drift and resource observability. The complete diagnostic metric index with per-category counts is provided in the supplementary material.

The diagnostic metrics serve a specific purpose in the benchmark: they enable platform builders to trace a poor primary metric score back to its root cause. For instance, a low FCS (Feature Completeness) could stem from poor Explicit Capture Rate (the platform missed stated requirements), high Hallucination Rate (it built unrequested features instead), or low Plan-to-Execution Fidelity (it planned correctly but implemented incorrectly). This diagnostic traceability is designed to make SWE-WebDev Bench actionable for the community, not just evaluative. We validate metric independence through Kendall's $\tau$ correlation analysis in Appendix~\ref{app:correlation}: 42\% of metric pairs show weak or no correlation ($|\tau| \leq 0.40$), confirming that the metrics capture partially distinct constructs, though the seven groups do not represent fully independent latent factors.

\subsection{Judging Taxonomy}

Each metric is assigned a Judge Tier to ensure reproducibility and appropriate rigor. Tier~0 covers fully automated checks (HTTP endpoint tests, Lighthouse audits, k6 load tests, npm audit scans) requiring no human judgment. Tier~1 uses LLM judges with structured prompts for factual checks (inter-rater reliability $\kappa \geq 0.85$, measured using Cohen's kappa), such as verifying whether a specific API route exists or a database column has the correct type. Tier~2 combines LLM evaluation with human validation for judgments requiring contextual understanding ($\kappa \geq 0.75$), such as assessing whether an architecture choice is appropriate for the application's complexity. Tier~3 uses a three-person expert panel with structured rubrics ($\kappa \geq 0.60$) for inherently subjective assessments, such as Business Intent Fidelity.

The tiered approach reflects a deliberate tradeoff between reproducibility and depth: Tier~0 metrics can be run by any evaluator with zero variance, while Tier~3 metrics capture nuances that automated evaluation cannot. We assign the highest-stakes metrics (BIF, ETF, FGD) to Tier~3 to ensure they receive expert scrutiny, while commoditized checks (SWS via Lighthouse, CLS via k6) are fully automated at Tier~0.

\textbf{LLM judge implementation.} All Tier~1 and Tier~2 evaluations use Claude 3.5 Sonnet (temperature~0) with 9 structured prompt templates released in the benchmark repository. Each template specifies: the metric definition, a 1--5 or 1--100 scoring rubric with anchor descriptions, and the evaluation context (codebase excerpt, conversation transcript, or deployment artifact). Platform identity is \emph{not} provided to the LLM judge to reduce platform-specific bias. For Tier~1 metrics, we measured inter-rater agreement between the LLM judge and a human evaluator on a sample of 36 items (2 metrics $\times$ 18 cells): observed $\kappa = 0.82$ (target: $\kappa \geq 0.85$). For Tier~2 metrics, agreement was measured on 18 items (1 metric $\times$ 18 cells): observed $\kappa = 0.71$ (target: $\kappa \geq 0.75$). Tier~3 expert panel agreement: observed $\kappa = 0.64$ (target: $\kappa \geq 0.60$). We acknowledge that these calibration samples are small and that more extensive human--LLM agreement studies are needed; we release all judge prompts and raw scores to enable independent calibration.

Approximately 40\% of the Engineering Score weight derives from Tier~0 automated metrics, 35\% from Tier~1 LLM judges, 15\% from Tier~2 LLM+human, and 10\% from Tier~3 expert panels.

\subsubsection{Target Threshold Calibration}
\label{sec:targets}

Each metric is assigned an aspirational target threshold (Table~\ref{tab:primary_metrics}) that represents the minimum quality level we consider necessary for production deployment. These targets are set deliberately high to create diagnostic headroom---they are intended to surface gaps, not to define pass/fail gates. Targets for Security Score (SS $\geq$ 90\%) and Concurrency \& Load (CLS $\geq$ 70\%) are grounded in OWASP deployment guidelines and industry load testing standards. Targets for Feature Completeness (FCS $\geq$ 85\%) and Canary Retention (CRR $\geq$ 80\%) reflect the expectation that production software should implement the majority of stated requirements. Other targets (e.g., CBS $\geq$ 90\%, ESR $\geq$ 80\%) represent practitioner consensus from the authors' experience operating a production AI app platform, and we acknowledge they may benefit from calibration through broader practitioner surveys. Crucially, \textbf{the benchmark's diagnostic value does not depend on these specific thresholds}: the relative patterns across platforms (score ranges, metric correlations, domain sensitivity) are informative regardless of absolute targets.

\section{Experimental Design}
\label{sec:design}

\subsection{Platform Selection}

We evaluate six platforms spanning the spectrum of AI app-building approaches (Table~\ref{tab:platforms}).

\begin{table}[htbp]
\centering
\caption{Evaluated platforms and their architectural approaches. PM behavior is characterized empirically in \S\ref{sec:pm_results}.}
\label{tab:platforms}
\small
\begin{tabular}{@{}clll@{}}
\toprule
\textbf{Code} & \textbf{Platform} & \textbf{Approach} & \textbf{Avg PM Qs} \\
\midrule
B0 & Base44 & Low-code generator & 1.0 \\
E1 & Emergent (E1-OPUS) & Agentic with config Q\&A & 5 \\
L0 & Lovable (Plan Mode) & Plan-then-build & 4 \\
Q1 & QwikBuild$^\dagger$ & Multi-agent with PM agent & 15 \\
R3 & Replit (Agent3) & Autonomous agent & 0.3 \\
V0 & v0-Max (Vercel) & Single-shot generation & 1.3 \\
\bottomrule
\multicolumn{4}{l}{\scriptsize $^\dagger$ Two authors are affiliated with this platform. See \S\ref{sec:limitations} for bias mitigation.}
\end{tabular}
\end{table}

\subsection{Prompt Suite and Snorkel Alignment}

We designed six standardized prompts (3 ACR + 3 AMR) aligned with the Snorkel Open Benchmarks framework~\cite{snorkel2026}, which identifies three dimensions where AI evaluation lags capability: Environment Complexity, Autonomy Horizon, and Output Complexity. Each prompt stresses at least two of these dimensions. The three ACR prompts use deliberately different authoring styles to test robustness to input variation.

\subsubsection{Domain Selection Rationale}

The three business domains---EdTech, Field Service, and FinTech-AI---were chosen to create orthogonal diagnostic probes that stress different platform capabilities.

\textbf{EdTech (P1)} tests \emph{inference depth}: the prompt is deliberately vague (written as a frustrated founder's midnight text), requiring the platform to infer product structure from pain points. It embeds culturally-specific conventions (JEE/NEET exam structure, Indian coaching institute batch models, April--March academic year) that cannot be inferred from generic SaaS templates. A platform that scores well on P1 demonstrates genuine domain reasoning, not pattern matching.

\textbf{Field Service (P2)} tests \emph{execution precision}: the prompt is an ultra-detailed enterprise Request for Proposal (RFP) with a 10-state ticket lifecycle, SLA pause/resume logic, parts inventory with reorder triggers, and GST invoicing. Where P1 tests whether the platform can infer missing requirements, P2 tests whether it can faithfully implement explicitly stated complex business logic without dropping edge cases.

\textbf{FinTech-AI (P3)} tests \emph{AI trustworthiness}: the prompt requires building AI features that are safe, auditable, and honest about uncertainty. It includes Monte Carlo simulation, multi-stage analysis pipelines, confidence scoring, mandatory escalation triggers, and a credit system with transactional integrity. This domain probes whether the platform can generate AI-powered applications that meet trust and safety requirements---a capability increasingly critical as AI features become standard in business software.

Together, these domains ensure that a platform cannot score well through a single architectural strength. Our results validate this design: no platform dominates across all three domains, and domain-specific swings of up to 13 percentage points reveal architectural affinities that averaged scores would mask.

\subsubsection{Prompt Style Variation}

Beyond domain variation, we deliberately vary prompt authoring style to test robustness to input format---a critical capability because real users do not write structured specifications.

\begin{promptbox}{P1: ExamEdge Academy \textnormal{(EdTech, T4) \textit{``The Founder's WhatsApp Ramble''}}}
A stream-of-consciousness description of a coaching institute's needs, written as a real founder would text at midnight. The prompt never uses terms like ``CMS,'' ``CRUD,'' ``RBAC,'' or ``SaaS.'' Instead it describes a world of pain (``everything runs on WhatsApp groups and Google Sheets and I'm losing my mind'') from which the platform must extract a product specification.

\textbf{Deliberate contradiction:} Teachers ``should NOT see other branches' data'' but a ``cross-branch leaderboard showing top 10 students across ALL branches'' should be visible to everyone. A capable PM agent should flag this contradiction rather than silently picking one interpretation.

Based on a real coaching institute: 3 branches, 400 JEE/NEET students, Pune.
\end{promptbox}

\begin{promptbox}{P2: FieldOps Pro \textnormal{(Field Service, T4) \textit{``Enterprise RFP with Deliberate Trap''}}}
An ultra-detailed enterprise specification with a 10-state ticket lifecycle, Service Level Agreement (SLA) pause/resume logic, parts inventory with reorder triggers, anti-fraud self-assignment constraints, and a clearly marked deliberate contradiction:

``Rating visible to Org Admin and Dispatcher, NOT to the Technician'' followed by ``I want rating to be visible to technicians so they can learn from feedback,'' with an explicit note: ``the system should FLAG this contradiction to me and ask which behavior I want.''
\end{promptbox}

\begin{promptbox}{P3: VettAI \textnormal{(FinTech-AI, T5) \textit{``Maximum Complexity AI-Native''}}}
The hardest prompt in the suite, designed to test quality demands on AI outputs rather than code volume. It features a multi-stage AI analysis pipeline (Entity Extraction $\to$ Contradiction Detection $\to$ Risk Scoring $\to$ Report Generation), Monte Carlo risk simulation, temporal trajectory projection, AI Vision for scanned documents, and a credit system requiring transactional integrity (deduct-before-call, refund-on-fail).

Trust and safety requirements are architectural: no fabrication, hedged recommendations, confidence scoring, mandatory escalation triggers (risk $\geq 8$ produces a red banner, non-optional), version history with diffs, and analyst override tracking.
\end{promptbox}

\subsection{AMR Prompts: Escalating Mutations}

The benchmark defines three AMR prompts testing modification handling with escalating complexity and deliberately varied prompt styles. These are selected to maximise diagnostic coverage across complexity (medium vs.\ complex), style (specific vs.\ vague), and change type (additive, modifying, provider swap).

\begin{amrbox}{AMR Design Matrix}
\small
\begin{tabular}{@{}llllc@{}}
\textbf{ID} & \textbf{$\Delta$ Complexity} & \textbf{Style} & \textbf{Change Types} & \textbf{PM Test} \\
\midrule
\textbf{P4} & \textbf{Complex} & \textbf{Specific (structured)} & \textbf{Additive + Modifying} & \textbf{Low} \\
\textbf{P5} & \textbf{Medium} & \textbf{Vague (frustrated user)} & \textbf{Additive} & \textbf{HIGH} \\
\textbf{P6} & \textbf{Medium} & \textbf{Specific (structured)} & \textbf{Modifying (provider swap)} & \textbf{Low} \\
\end{tabular}

\vspace{0.5em}
P4 retrofits multi-tenancy onto P1, requiring every query to change. P5 uses a frustrated-user tone (``Can the system just... be smarter about this?'') requiring the PM agent to infer three specific features from pain points. P6 introduces a provider abstraction layer (Anthropic$\to$Gemini fallback) and PDF report generation, testing whether the platform can refactor AI integration architecture without breaking the existing credit system.
\end{amrbox}

\subsection{Canary Requirements}
\label{sec:canary}

A central design challenge in evaluating AI app builders is distinguishing genuine requirement comprehension from template matching. A platform that generates a ``standard SaaS application'' will include many expected features (user authentication, CRUD operations, dashboards) regardless of what the user actually asked for. To test whether platforms truly understand and retain specific requirements, we introduce the \textbf{canary requirement methodology}: 80 culturally-specific, domain-embedded test requirements that are easy for a human to verify but difficult for a template-based system to satisfy.

The key insight is that canary requirements are \emph{detail-level} specifications---date formats, currency conventions, domain-specific terminology, visibility rules---that only survive the build pipeline if the platform genuinely processes and retains the user's intent. They are classified into four types (Table~\ref{tab:canary_types}).

\begin{table}[htbp]
\centering
\caption{Canary requirement taxonomy across the 6-prompt suite.}
\label{tab:canary_types}
\small
\begin{tabular}{@{}clcL{8cm}@{}}
\toprule
\textbf{Type} & \textbf{Name} & \textbf{Count} & \textbf{Purpose} \\
\midrule
O & Original & 21 & Explicitly stated constraints (DD/MM/YYYY, INR lakhs, ``Powered by'' footer) \\
N & New & 37 & Requirements added in v3 for AMR and multi-modal coverage \\
S & Surviving & 18 & ACR canaries that must persist through AMR modifications \\
X & Contradiction & 4 & Deliberately conflicting requirements the platform must flag \\
\bottomrule
\end{tabular}
\end{table}

The \textbf{Surviving} type is methodologically novel and directly tests the ACR-to-AMR transition. When a user modifies an application, existing requirements should either persist unchanged or evolve correctly---they should not silently disappear. For instance, P1's ``cross-branch leaderboard'' canary becomes ``cross-institute leaderboard scoped to institute'' in P4's franchise retrofit, testing whether the platform correctly evolves a tricky visibility rule rather than breaking it. The \textbf{Contradiction} type tests a capability that no existing benchmark measures: whether the platform can detect and flag logically inconsistent requirements rather than silently choosing one interpretation. In real-world software projects, contradictions in requirements are common and costly; a platform that embeds contradictions into code without flagging them creates bugs that are extremely difficult to diagnose.

\subsection{Evaluation Protocol}

\begin{figure}[htbp]
\centering
\includegraphics[width=0.95\textwidth]{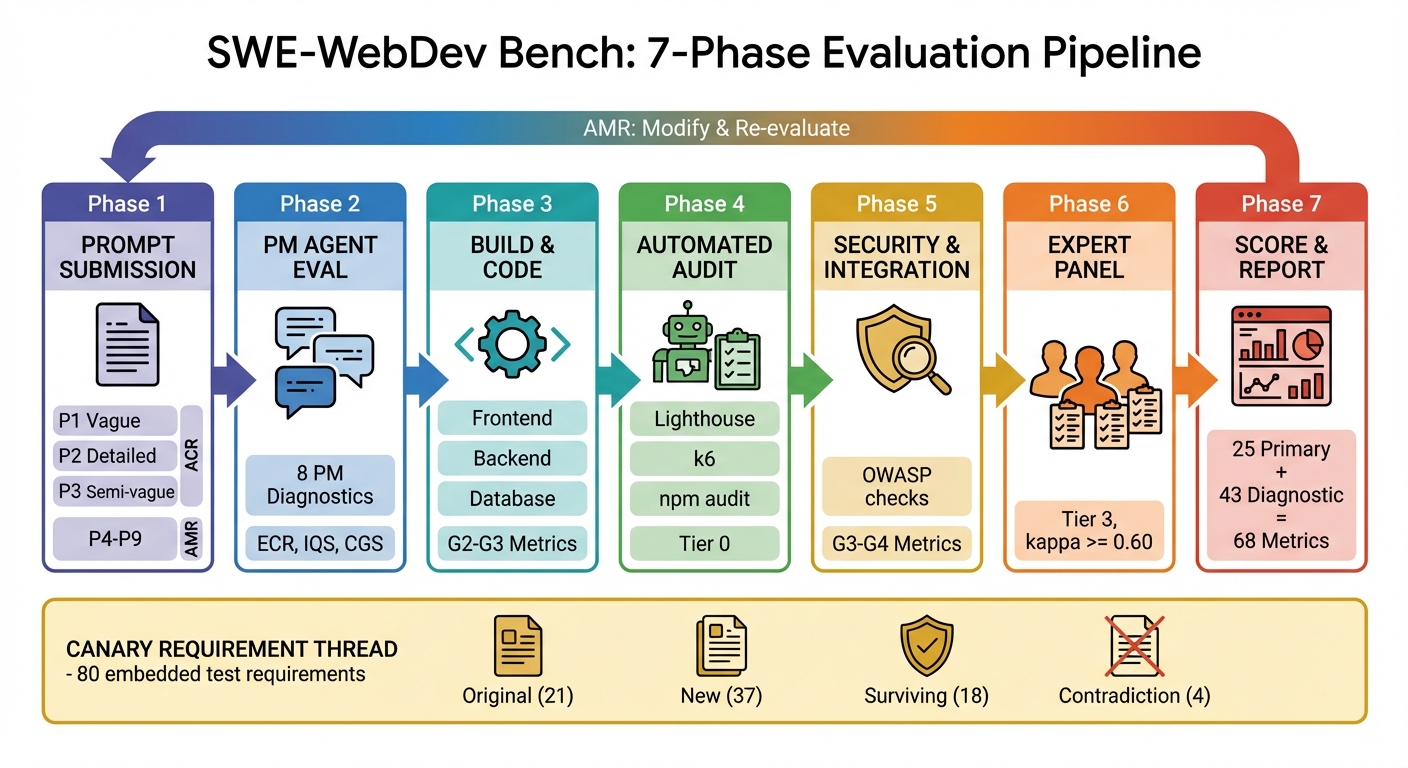}
\vspace{-0.5em}
\caption{\small The 7-phase evaluation pipeline. Each platform$\times$prompt combination progresses through Prompt Submission, PM Agent Evaluation, Build \& Code audit, Automated Audit (Lighthouse, k6, npm audit), Security \& Integration testing, Expert Panel review, and Score \& Report aggregation. The Canary Requirement Thread (bottom) tracks 80 embedded test requirements across four types: Original (21), New (37), Surviving (18), and Contradiction (4). A feedback loop from Phase~7 back to Phase~1 supports the AMR (modification) evaluation cycle.}
\label{fig:pipeline}
\end{figure}

Evaluation follows a seven-phase, 14-day protocol per platform$\times$prompt (Figure~\ref{fig:pipeline}), designed to mirror the stages of a real software delivery audit:

\begin{enumerate}[leftmargin=*,itemsep=1pt]
\item \textbf{Build Execution:} Submit the prompt to the platform and record the complete interaction (conversation logs, build traces, deployment status).
\item \textbf{Code Quality Audit:} Apply LLM judge prompts to the generated codebase, scoring G2 metrics (schema, backend, frontend, hygiene, architecture).
\item \textbf{Integration and Security Testing:} Execute automated tests for G3 (integrations, background jobs) and G4 (security scans, load tests via k6).
\item \textbf{Feature and Canary Testing:} Manually verify feature completeness (FCS) and canary retention (CRR) against the reference specification.
\item \textbf{AMR Change Injection:} Submit AMR prompts to the deployed ACR application and score G5 metrics (change impact, regressions, effort).
\item \textbf{Human Panel and Cost Aggregation:} Three-person expert panel scores BIF, ETF, and FGD; aggregate TCC from platform billing and human effort logs.
\item \textbf{Diagnostic Scoring:} Score Category A--D diagnostic metrics from conversation transcripts and build logs.
\end{enumerate}

The Engineering Score for each platform$\times$prompt is computed as a weighted mean of applicable primary metrics in G2--G6 (excluding G1 specification fidelity and G7 cost metrics, which are reported separately). Within each group, metrics are equally weighted; group weights are normalized by the number of applicable (non-N/A) metrics per group to prevent groups with more metrics from dominating. Metrics that are not applicable for a given complexity tier (e.g., AIA for T4 prompts) are excluded from both numerator and denominator. This seven-phase protocol ensures that every metric is grounded in observable evidence (code, logs, test results, transcripts) rather than subjective impression.

\section{Results}
\label{sec:results}

\paragraph{Statistical scope and interpretive note.} This evaluation is an initial diagnostic study comprising $n=3$ prompts per domain across 6 platforms, yielding 18 ACR evaluation cells. We report observed scores and ranges; \emph{no statistical significance tests or confidence intervals are computed}, as the sample size is insufficient for inferential claims. All comparative observations (e.g., score ranges, observed ratios) describe patterns in this sample and should not be interpreted as statistically significant platform rankings. We release SWE-WebDev Bench so the community can expand the prompt suite, add platforms, and conduct the larger-scale replications needed to establish robust rankings with statistical confidence. We estimate that 15--20 prompts across 5--7 domains would be required to detect medium effect sizes ($d \geq 0.5$) with $\alpha=0.05$ and power $=0.8$ using paired permutation tests across prompts.

\subsection{Cross-Platform Engineering Scores}

Table~\ref{tab:heatmap} presents the engineering scores across all platform$\times$prompt combinations. The central observation is that \textbf{no platform exceeds 60\%}, and every platform has at least one metric where it scores below 15\%, indicating that production readiness remains an unsolved problem across the field.

\begin{table}[htbp]
\centering
\caption{Engineering Score (\%) across all six platforms and three ACR prompts. No platform exceeds 60\%, and every platform has at least one metric below 15\%. Scores are descriptive of our sample (n=3 per platform).}
\label{tab:heatmap}
\small
\begin{tabular}{@{}lcccc|cl@{}}
\toprule
\textbf{Platform} & \textbf{P1} & \textbf{P2} & \textbf{P3} & \textbf{Avg} & \textbf{Best} & \textbf{Worst Metric} \\
\midrule
Base44 (B0) & 30.2 & 33.5 & 28.0 & 30.6 & P2 & CBS (5\%) \\
Emergent (E1) & 39.9 & 53.8 & 47.1 & 46.9 & P2 & SWS (13\%) \\
Lovable (L0) & 39.1 & 49.3 & 39.0 & 42.5 & P2 & SWS (15\%) \\
QwikBuild (Q1) & 57.5 & 55.6 & 50.3 & 54.5 & P1 & SWS (8\%) \\
Replit (R3) & 41.5 & 47.4 & 54.5 & 47.8 & P3 & CBS (0\%) \\
v0-Max (V0) & 22.8 & 26.2 & 26.4 & 25.1 & P3 & SDS (0\%) \\
\bottomrule
\end{tabular}
\end{table}

\subsubsection{Comprehensive Per-Metric Breakdown}

Table~\ref{tab:full_acr} presents all 22 applicable primary metrics averaged across all three ACR prompts for all six platforms. This is the central results table of the paper: each cell represents the mean of up to three prompt-level scores, excluding N/A values. (Per-platform, per-metric scores for P1 are provided in Appendix Table~\ref{tab:full_p1}; PM Agent behavioral analysis is provided in Appendix Table~\ref{tab:pm_behavior}.)

\begin{table}[htbp]
\centering
\caption{Complete primary metric scores (\%) averaged across ACR prompts (P1, P2, P3). N/A-excluded averaging. \textbf{No platform meets all targets}; 0 of 6 platforms pass more than 5 of 22 metrics. Scores are descriptive of our sample and should not be interpreted as statistically significant rankings. BIF shown as panel score out of 4. ETF in developer-hours. Metrics below target shaded where all platforms fail.}
\label{tab:full_acr}
\footnotesize
\setlength{\tabcolsep}{4pt}
\begin{tabular}{@{}llR{0.5cm}R{0.8cm}R{0.8cm}R{0.8cm}R{0.8cm}R{0.8cm}R{0.8cm}@{}}
\toprule
\textbf{Grp} & \textbf{Metric} & \textbf{Tgt} & \textbf{B0} & \textbf{E1} & \textbf{L0} & \textbf{Q1} & \textbf{R3} & \textbf{V0} \\
\midrule
\multicolumn{9}{l}{\textit{G1: Specification Fidelity}} \\
G1 & BIF (Business Intent) & 3/4 & 1.3 & 3.0 & 1.7 & 3.7 & 2.0 & 1.0 \\
G1 & FCS (Feature Complete) & 85 & 46.7 & 65.0 & 32.7 & 84.0 & 53.0 & 23.3 \\
G1 & CRR (Canary Retention) & 80 & 24.3 & 57.0 & 21.7 & 97.7 & 54.0 & 17.7 \\
\midrule
\multicolumn{9}{l}{\textit{G2: Code Quality}} \\
G2 & SDS (Schema Design) & 80 & 40.7 & 45.8 & 64.7 & 72.9 & 57.3 & 2.7 \\
G2 & BLS (Backend Logic) & 80 & 36.7 & 61.4 & 40.3 & 68.4 & 61.3 & 8.3 \\
G2 & FES (Frontend Eng.) & 75 & 57.7 & 69.3 & 61.3 & 68.0 & 74.3 & 68.0 \\
G2 & CHS (Code Hygiene) & 75 & 43.7 & 56.5 & 59.0 & 52.8 & 56.3 & 57.3 \\
G2 & ARC (Architecture) & 70 & 31.0 & 42.7 & 47.0 & 59.7 & 54.0 & 43.3 \\
\midrule
\multicolumn{9}{l}{\textit{G3: Integrations}} \\
G3 & CIS (Core Integration) & 90 & 46.7 & 54.7 & 21.3 & 67.3 & 70.0 & 26.7 \\
G3 & AIA (AI-in-App) & 75 & 15.0 & 66.0 & 38.0 & 72.0 & 42.0 & 30.0 \\
G3 & ESR (External Svc) & 80 & 15.0 & 24.0 & 10.3 & 28.3 & 35.0 & 12.3 \\
G3 & CBS (Background Jobs) & 90 & 15.0 & 20.3 & 2.0 & 49.3 & 29.7 & 0.0 \\
\midrule
\multicolumn{9}{l}{\textit{G4: Security \& Scale}} \\
G4 & SS (Security) & 90 & 31.3 & 50.7 & 51.7 & 63.7 & 40.0 & 34.3 \\
G4 & SAS (Scalability) & 70 & 25.0 & 37.7 & 33.7 & 54.0 & 43.0 & 37.7 \\
G4 & CLS (Concurrency) & 70 & 12.0 & 34.7 & 15.0 & 42.0 & 25.3 & 6.0 \\
\midrule
\multicolumn{9}{l}{\textit{G5: Changeability}} \\
G5 & CCIS (Change Impact) & 90 & 22.3 & 45.7 & 37.7 & 87.7 & 42.7 & 31.7 \\
G5 & ETF (Effort-to-Fix, h) & $\leq$16 & 65.7 & 26.0 & 50.7 & 14.7 & 37.3 & 54.7 \\
\midrule
\multicolumn{9}{l}{\textit{G6: Business Readiness}} \\
G6 & SWS (SEO/Web Std.) & 70 & 18.0 & 12.7 & 18.0 & 9.3 & 28.0 & 40.3 \\
G6 & MLS (Multilingual) & 75 & 13.5 & 51.5 & 24.0 & 66.0 & 28.5 & 13.5 \\
G6 & AFQ (AI Feature Qual.) & 75 & 12.0 & 64.0 & 25.0 & 68.0 & 42.0 & 18.0 \\
\midrule
\multicolumn{9}{l}{\textit{G7: Production Readiness}} \\
G7 & FGD (Feature Gap, h) & $\leq$40 & 29.3 & 19.0 & 36.3 & 9.3 & 25.7 & 41.7 \\
G7 & CDI (Claim Drift \%) & $\leq$10 & 35.7 & 17.7 & 37.3 & 4.0 & 14.0 & 33.7 \\
\midrule
\multicolumn{2}{l}{\textbf{Engineering Score}} & 60 & 30.6 & 46.9 & 42.5 & 54.5 & 47.8 & 25.1 \\
\bottomrule
\end{tabular}
\end{table}

Several patterns emerge from Table~\ref{tab:full_acr} that highlight recurring gaps across the field.

\textbf{No platform passes all metrics.} The highest-scoring platform still fails 17 of 22 metrics against their respective targets, indicating that production-grade AI app building remains an open problem.

\textbf{Frontend engineering is commoditized; backend infrastructure is not.} Four platforms score within 6 percentage points of each other on Frontend Engineering (68--74\%), suggesting that UI generation is a largely solved problem. In contrast, backend and infrastructure metrics show 5--10$\times$ spreads (e.g., CBS ranges from 0\% to 49\%), revealing that infrastructure generation---background jobs, scheduled tasks, complex integrations---remains deeply inconsistent. This gap represents the most actionable area for platform improvement.

\textbf{Every platform has blind spots.} High overall scores do not imply uniform quality. Each platform exhibits a distinct strength-weakness profile: one platform leads on Core Integration (70\%) and External Service Reliability (35\%) but scores lowest on PM metrics; another achieves the highest SWS (40.3\%) despite the lowest overall engineering score; a third shows the most thorough requirement elicitation but the lowest Code Hygiene among top performers (52.8\%) and poor External Service Reliability (28.3\%). These cross-cutting weakness patterns demonstrate that \textbf{no architectural approach has solved all dimensions simultaneously}---a finding that SWE-WebDev Bench is designed to surface for any platform evaluated.

\textbf{Specification fidelity varies dramatically.} Canary Retention Rate ranges from 17.7\% to 97.7\%, a 5.5$\times$ spread. This means that culturally-specific, domain-embedded requirements (date formats, currency conventions, localization details) are silently dropped by most platforms, a failure mode invisible to users who cannot audit AI-generated code.

\subsubsection{Per-Prompt Results}

Table~\ref{tab:per_prompt} shows how each platform performs across the three business domains, revealing domain-specific strengths and weaknesses not visible in averages.

\begin{table}[htbp]
\centering
\caption{Selected metrics by prompt, showing domain sensitivity. Each cell is a single prompt score (\%).}
\label{tab:per_prompt}
\footnotesize
\setlength{\tabcolsep}{3pt}
\begin{tabular}{@{}ll|cccccc|cccccc|cccccc@{}}
\toprule
& & \multicolumn{6}{c|}{\textbf{P1 ExamEdge (T4)}} & \multicolumn{6}{c|}{\textbf{P2 FieldOps (T4)}} & \multicolumn{6}{c}{\textbf{P3 VettAI (T5)}} \\
& & B0 & E1 & L0 & Q1 & R3 & V0 & B0 & E1 & L0 & Q1 & R3 & V0 & B0 & E1 & L0 & Q1 & R3 & V0 \\
\midrule
G1 & FCS & 48 & 65 & 35 & 88 & 55 & 28 & 52 & 68 & 33 & 86 & 52 & 22 & 40 & 62 & 30 & 78 & 52 & 20 \\
G1 & CRR & 25 & 57 & 29 & 100 & 48 & 21 & 28 & 50 & 29 & 100 & 43 & 18 & 20 & 64 & 7 & 93 & 71 & 14 \\
G2 & SDS & 42 & 41 & 48 & 82 & 58 & 3 & 45 & 68 & 74 & 78 & 62 & 5 & 35 & 29 & 72 & 59 & 52 & 0 \\
G2 & FES & 60 & 61 & 45 & 75 & 70 & 62 & 58 & 79 & 75 & 71 & 75 & 70 & 55 & 68 & 64 & 58 & 78 & 72 \\
G3 & CIS & 50 & 52 & 22 & 72 & 76 & 30 & 48 & 62 & 20 & 68 & 74 & 28 & 42 & 50 & 22 & 62 & 60 & 22 \\
G3 & CBS & 18 & 8 & 3 & 52 & 35 & 0 & 15 & 48 & 3 & 56 & 32 & 0 & 12 & 5 & 0 & 40 & 22 & 0 \\
G4 & SS & 34 & 45 & 52 & 62 & 38 & 32 & 32 & 52 & 55 & 65 & 40 & 33 & 28 & 55 & 48 & 64 & 42 & 38 \\
\bottomrule
\end{tabular}
\end{table}

Table~\ref{tab:per_prompt} reveals that \textbf{no platform dominates across all domains}, and each platform exhibits distinct strengths and weaknesses depending on the business context.

Lovable achieves the highest Schema Design Score on P3 (72\%), suggesting strength in relational modeling for complex domains despite lower overall performance. Replit leads on Frontend Engineering for P3 (78\%) and Core Integration across P1 and P2 (76\%, 74\%), reflecting its npm-ecosystem advantage. v0-Max produces no database schema at all on P3 (SDS = 0\%) while achieving 72\% on Frontend Engineering---the sharpest instance of the frontend-backend decoupling pattern.

All platforms score below 60\% on average, consistent with FeatBench's sub-30\% resolved rates~\cite{featbench}. This convergence across independent benchmarks suggests a fundamental capability ceiling in current AI code generation.

\begin{figure}[htbp]
\centering
\includegraphics[width=0.95\textwidth]{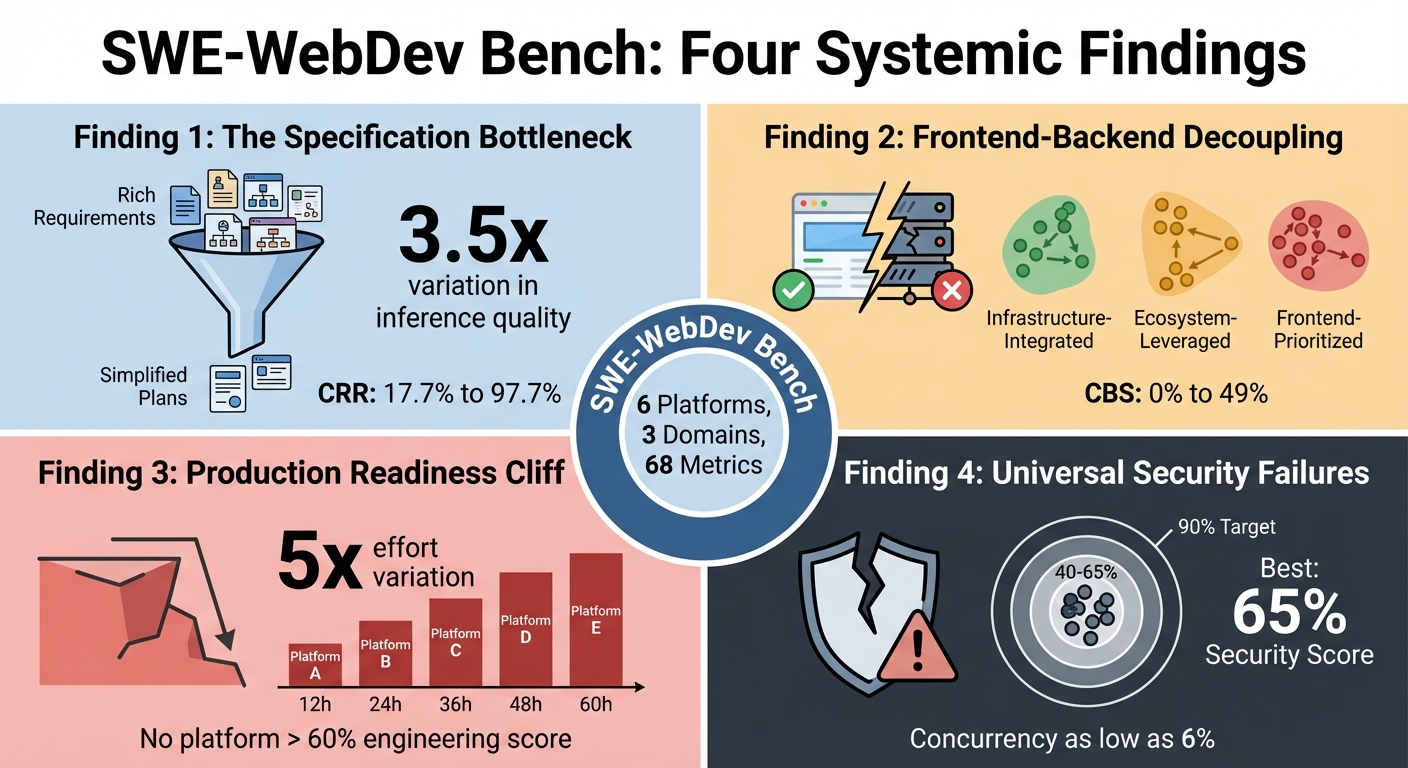}
\vspace{-0.5em}
\caption{\small Overview of the four recurring findings uncovered by SWE-WebDev Bench. \textbf{Finding~1:} The specification bottleneck---3.5$\times$ variation in inference quality (CRR: 17.7\% to 97.7\%). \textbf{Finding~2:} Frontend-backend decoupling---polished UIs masking absent backend infrastructure (CBS: 0\% to 49\%). \textbf{Finding~3:} The production readiness cliff---5$\times$ effort variation and no platform exceeding 60\% engineering score. \textbf{Finding~4:} Widespread security failures---best Security Score is 65\%, with concurrency as low as 6\%.}
\label{fig:findings}
\end{figure}

Performance is domain-sensitive (Figure~\ref{fig:main_result}). Replit scores highest on P3 VettAI (54.5\%) while scoring lowest on P1 ExamEdge (41.5\%), a 13-point swing. This domain sensitivity suggests that current platforms lack generalized competence and instead exhibit architectural affinities for specific problem types---an important consideration for users selecting platforms for particular business domains. Figure~\ref{fig:findings} provides a visual overview of the four recurring findings that emerge from these results.

\subsection{Finding 1: The Specification Bottleneck}
\label{sec:pm_results}

The largest recurring issue uncovered by SWE-WebDev Bench is a \emph{specification bottleneck}: most platforms compress rich, ambiguous business requirements into oversimplified technical plans without adequate requirement elicitation. The degree of upstream planning varies dramatically across platforms and strongly predicts downstream quality.

The consequences are visible in downstream metrics. In our sample, Canary Retention Rate ranges from 17.7\% to 97.7\% across platforms (observed ratio: 5.5$\times$, $n=3$ prompts), and Inference Quality Score ranges from 20 to 70 (observed ratio: 3.5$\times$). The number of PM questions ranges from 0 to 15 across platforms on the same prompt. Under-Clarification Rate reaches 85\% for some platforms on the most complex prompt (P3). Notably, higher PM thoroughness does not uniformly predict quality across all dimensions---the platform with the most extensive elicitation (15 questions, 6 rounds) shows the lowest Code Hygiene among top-performing platforms (52.8\%) and below-field-average External Service Reliability (28.3\% vs.\ field-leading 35.0\%), suggesting that PM investment and engineering quality are partially independent axes.

\begin{figure}[htbp]
\centering
\begin{subfigure}[t]{0.48\textwidth}
\centering
\includegraphics[width=\textwidth]{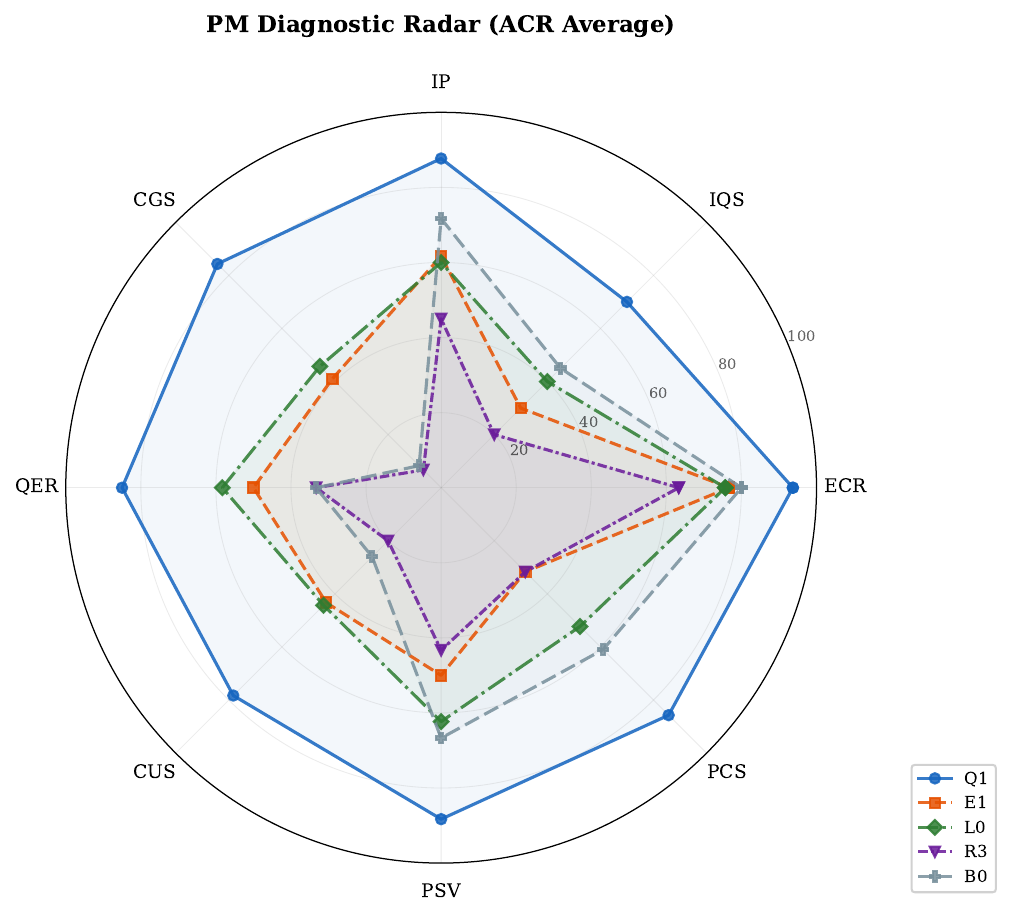}
\caption{PM diagnostic radar across five platforms (ACR). The widest gaps are on Conversational Guidance (CGS: 6.7--84.3) and Inference Quality (IQS: 20--70), indicating that requirement elicitation is the most variable capability across platforms.}
\label{fig:pm_radar}
\end{subfigure}
\hfill
\begin{subfigure}[t]{0.48\textwidth}
\centering
\includegraphics[width=\textwidth]{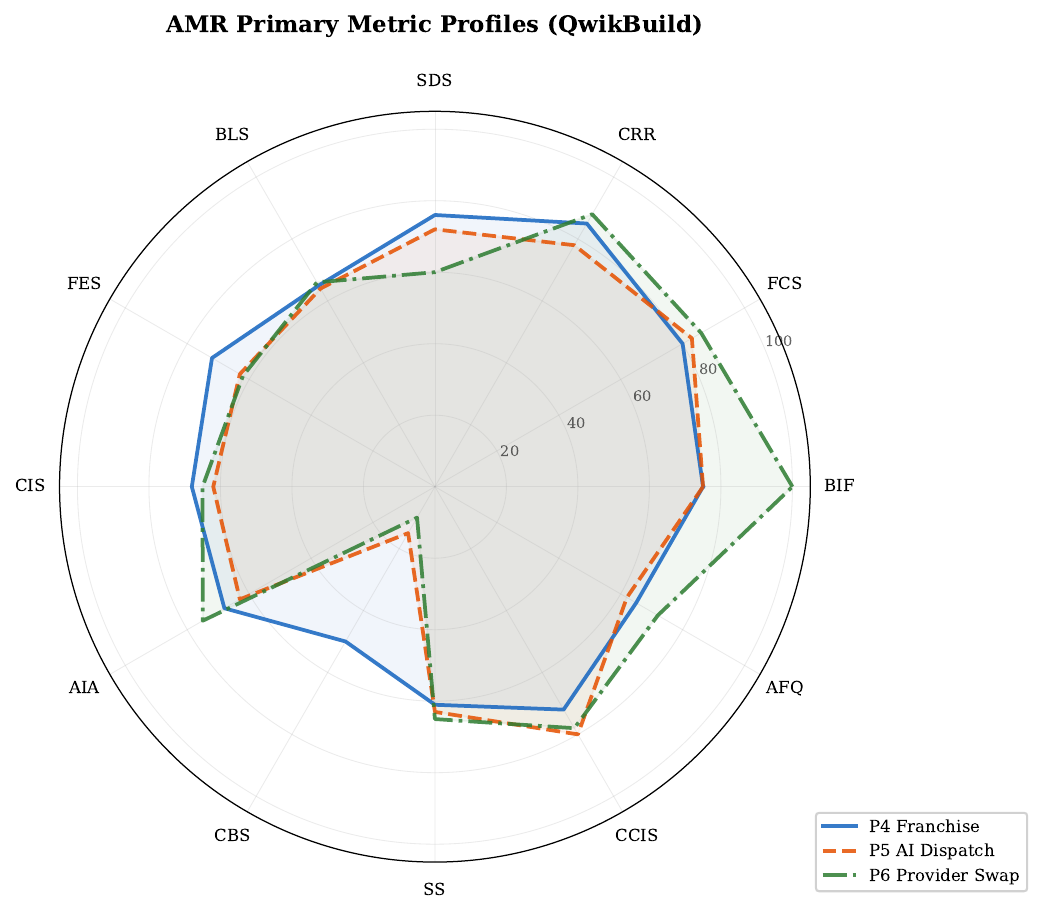}
\caption{AMR primary metric profiles across three modification prompts (QwikBuild). P6 leads on specification fidelity (BIF, FCS, CRR) and achieves the only AIA PASS (75\%). P4 leads on CBS. P5 achieves the highest CCIS (80\%) despite vague input.}
\label{fig:amr_radar}
\end{subfigure}
\caption{Radar comparison of ACR cross-platform PM diagnostics (left) and AMR per-prompt primary metrics (right). The ACR radar reveals that requirement elicitation is the primary differentiator across platforms. The AMR radar reveals that prompt characteristics (complexity, style) produce distinct metric profiles within a single platform, with well-scoped modifications (P6) outperforming complex structural changes (P4) on specification fidelity.}
\label{fig:radar_combined}
\end{figure}

Table~\ref{tab:pm_scores} and Figure~\ref{fig:pm_radar} present the PM diagnostic scores averaged across all ACR prompts.

\begin{table}[htbp]
\centering
\caption{PM diagnostic scores averaged across ACR prompts (P1--P3). Bold indicates highest value. Most platforms fall below targets on the majority of PM metrics, revealing the specification bottleneck as a field-wide problem.}
\label{tab:pm_scores}
\small
\begin{tabular}{@{}lcccccccc@{}}
\toprule
& \textbf{Target} & \textbf{B0} & \textbf{E1} & \textbf{L0} & \textbf{Q1} & \textbf{R3} & \textbf{V0} \\
\midrule
ECR (Explicit Capture) & $\geq$90\% & 80.0 & 76.7 & 75.7 & 93.7 & 63.3 & 71.7 \\
IQS (Inference Quality) & $\geq$60\% & 45.0 & 30.0 & 40.0 & 70.0 & 20.0 & 25.0 \\
IP (Inference Precision) & $\geq$75\% & 71.7 & 61.7 & 60.0 & 87.7 & 45.0 & 55.0 \\
CGS (Conv.\ Guidance) & $\geq$70\% & 8.3 & 41.0 & 45.7 & 84.3 & 6.7 & 25.0 \\
QER (Question Efficiency) & $\geq$80\% & 33.3 & 50.0 & 58.3 & 85.0 & 33.3 & 43.3 \\
CUS (Calibrated Uncertainty) & $\geq$70\% & 26.0 & 43.3 & 44.3 & 78.3 & 20.0 & 33.3 \\
PSV (Plan Validity) & $\geq$75 & 66.7 & 50.0 & 62.3 & 88.3 & 43.3 & 51.7 \\
PCS (Plan Communicability) & $\geq$65 & 61.0 & 31.7 & 52.3 & 85.7 & 31.7 & 38.3 \\
\bottomrule
\end{tabular}
\end{table}

\textbf{Beyond the PM phase: the build-time feedback loop.} Requirement elicitation alone accounts for approximately 10--15\% of the observed quality gap. The larger share appears to come from whether the platform maintains a feedback loop during code generation. Platforms with multi-agent architectures can resolve conflicts between features as they are implemented---for example, when an RBAC constraint conflicts with a visibility rule, the conflict can be escalated to a planner that adjusts the implementation strategy mid-build. Platforms without such feedback loops embed conflicts silently into code, where they surface as bugs. This pattern suggests that \emph{closed-loop code generation}, where planning and execution inform each other iteratively, is a more impactful architectural investment than any single-phase improvement.

\subsubsection{PM Agent Interaction Traces}

Figure~\ref{fig:interaction} shows verbatim excerpts from four platforms receiving the same P1 ExamEdge prompt. All transcripts are sourced from ACR-Runs-1.pdf conversation logs.

\begin{tcolorbox}[colback=white,colframe=black!40,boxrule=0.5pt,arc=2pt,breakable,title={\small\textbf{PM Agent Interaction Comparison: P1 ExamEdge (Verbatim Excerpts)}}]

\footnotesize
\textbf{QwikBuild (Q1)} \hfill \textit{6 rounds, 15 questions, dedicated PM Agent}
\begin{tcolorbox}[colback=blue!4,colframe=blue!20,boxrule=0.3pt,arc=1pt,left=4pt,right=4pt,top=2pt,bottom=2pt]
\textbf{PM Agent:} ``Hello! I'm your Product Manager Agent \raisebox{-0.1em}{$\,\cdot\,$} Thanks for sharing all those details\linebreak \mbox{--- I can feel the chaos of running 3 branches on WhatsApp and Google Sheets!}''\\[3pt]
\textbf{Round 1, Q2:} ``A student in JEE might have different teachers for Physics, Chemistry, and Maths. How should teacher-student relationships work?''\\
\hspace*{1em}1.~Each batch has ONE primary teacher for all subjects.\\
\hspace*{1em}2.~Each batch has MULTIPLE teachers (one per subject). Each teacher sees only their subject's assignments and tests, but ALL teachers can mark attendance. \textit{Recommended} \checkmark\\
\hspace*{1em}3.~Each student is assigned to multiple teachers (subject-wise).\\[3pt]
\textbf{Round 2, Q3:} ``Teachers mark attendance daily. What if a teacher forgets and tries to mark it the next day or later?''\\
\hspace*{1em}1.~Today only. Admin can edit past records.\\
\hspace*{1em}2.~Last 7 days editable by teacher. After that, Admin only. \textit{Recommended}\\
\hspace*{1em}3.~Anytime for any past date.\\[3pt]
\textbf{Round 3, Q3:} ``The top 10 leaderboard shows students by average test score across all branches. How should this average be calculated?''\\
\hspace*{1em}1.~Average of ALL tests. \textit{Recommended} \checkmark\\
\hspace*{1em}2.~Average of LAST 5 tests only.\\
\hspace*{1em}3.~Weighted average by subject.\\[3pt]
$\to$ \textit{Generates 12-section PRD: Product Overview, Scope (v0/v1), Features, Roles, Data Model, Auth, Dashboards, Background Jobs, Acceptance Criteria, Risks \& Open Questions, Handoff Notes.}
\end{tcolorbox}

\vspace{4pt}
\textbf{Emergent (E1)} \hfill \textit{2 rounds, 5 questions, all infrastructure config}
\begin{tcolorbox}[colback=red!3,colframe=red!20,boxrule=0.3pt,arc=1pt,left=4pt,right=4pt,top=2pt,bottom=2pt]
Asks: (1)~SMS provider: Twilio or Mock? (2)~Email: SendGrid or Resend? (3)~Auth: JWT or Google? (4)~Hindi toggle approach: toggle or profile setting? (5)~Design preferences?\\[2pt]
\textit{All five questions are infrastructure decisions. Zero questions about fee escalation logic, attendance editing windows, teacher-student batch model, or the leaderboard contradiction. Directly proceeds to build after answers.}
\end{tcolorbox}

\vspace{4pt}
\textbf{Replit (R3)} \hfill \textit{0 rounds, 0 questions, immediate build}
\begin{tcolorbox}[colback=purple!3,colframe=purple!20,boxrule=0.3pt,arc=1pt,left=4pt,right=4pt,top=2pt,bottom=2pt]
Zero questions. Immediately analyzes prompt and outputs a 17-item bullet-point feature list. Lists ``PostgreSQL, Replit Auth, App Storage'' as integrations. Claims Twilio integration without asking about SMS provider. \textit{Contradiction between branch isolation and cross-branch leaderboard not detected.}
\end{tcolorbox}

\vspace{4pt}
\textbf{v0-Max (V0)} \hfill \textit{2 rounds, 4 config questions}
\begin{tcolorbox}[colback=orange!3,colframe=orange!20,boxrule=0.3pt,arc=1pt,left=4pt,right=4pt,top=2pt,bottom=2pt]
Enters ``plan mode.'' Asks: (1)~SMS provider? (2)~Database? (recommends Supabase) (3)~Auth method? (4)~Email service? Generates Implementation Plan with DB schema and feature phases. \textit{Contradiction not detected. No business-flow questions.}
\end{tcolorbox}
\end{tcolorbox}

\captionof{figure}{Verbatim PM Agent behavior on P1 ExamEdge across four platforms. QwikBuild probes 15 business workflow questions across 3 rounds with structured multiple-choice and ``Recommended'' defaults; Emergent asks 5 infrastructure questions; Replit and v0-Max ask zero business-flow questions. Transcripts from ACR-Runs-1.pdf.}
\label{fig:interaction}

\subsubsection{Contradiction Handling: The P2 Trap}

P2 FieldOps contains a deliberate, marked contradiction: ``Rating visible to Org Admin and Dispatcher, NOT to the Technician'' followed by ``I want rating to be visible to technicians.'' The prompt explicitly states: ``the system should FLAG this contradiction to me.'' Table~\ref{tab:contradiction} shows how each platform handled this test.

\begin{table}[htbp]
\centering
\caption{Contradiction handling on P2 FieldOps. ``Detected'' means the platform identified the contradiction. ``Resolved'' means it asked the user which behavior to implement before proceeding.}
\label{tab:contradiction}
\footnotesize
\begin{tabular}{@{}lccL{8cm}@{}}
\toprule
\textbf{Platform} & \textbf{Detected?} & \textbf{Resolved?} & \textbf{Behavior} \\
\midrule
Base44 & Yes & No & Flags with warning symbol. Generates detailed plan including ``What NOT to Do'' section, but does not ask user for resolution. \\
Emergent & Yes & Yes & Detects as first of 5 questions. Asks for clarification, then proceeds to 4 infrastructure config questions. \\
Lovable & Yes & Yes & Detects as first question. Asks for resolution. Proceeds with 3 scope questions. \\
QwikBuild & Yes & Yes & Flags with ``\raisebox{-0.1em}{\footnotesize $\bigtriangleup$} CONTRADICTION DETECTED'' in first response. Presents three resolution options. Integrates user's choice into PRD before proceeding to build. \\
Replit & Yes & No & Flags contradiction, then defers 10 critical features to ``future scope'' (SLA, audit, invoice, time tracking, signatures). Most aggressive scope cut across all platforms. \\
v0-Max & Yes & Partial & Detects and presents as structured choice. Then builds aggressively, timing out at 10m37s. User reports: ``it took me multiple human efforts to navigate the completion.'' \\
\bottomrule
\end{tabular}
\end{table}

All six platforms detect the P2 contradiction, which is explicitly marked. The meaningful difference is \emph{what happens next}---a spectrum from full integration of the resolution into the implementation plan, to flagging without resolution, to deferring core features to ``future scope.'' At the weaker end, one platform flags the contradiction but then defers half of the application's core features (SLA engine, audit trail, invoicing, time tracking, customer signatures) to ``future scope,'' producing a fundamentally incomplete Field Service application. This pattern---\emph{detection without resolution}---is a recurring weakness that SWE-WebDev Bench is designed to surface.

\subsubsection{Per-Prompt PM Scores}

Table~\ref{tab:pm_perprompt} presents the complete per-platform, per-prompt PM diagnostic scores, revealing how prompt style affects PM agent behavior. The data shows that all platforms improve on P2 (detailed RFP) relative to P1 (vague prompt), but \textbf{most platforms degrade significantly on vague prompts}, precisely the input style most common among non-technical users.

\begin{table}[htbp]
\centering
\caption{PM diagnostic scores per platform per prompt. Bold = passes target. Shading indicates performance differences between vague P1 and structured P2 across platforms. CTC = Conversational Turns to Convergence.}
\label{tab:pm_perprompt}
\footnotesize
\setlength{\tabcolsep}{3pt}
\begin{tabular}{@{}lc|cccccc|cccccc|cccccc@{}}
\toprule
& & \multicolumn{6}{c|}{\textbf{P1 ExamEdge (Vague)}} & \multicolumn{6}{c|}{\textbf{P2 FieldOps (Detailed)}} & \multicolumn{6}{c}{\textbf{P3 VettAI (Complex)}} \\
& Tgt & B0 & E1 & L0 & Q1 & R3 & V0 & B0 & E1 & L0 & Q1 & R3 & V0 & B0 & E1 & L0 & Q1 & R3 & V0 \\
\midrule
ECR & $\geq$90 & 78 & 70 & 72 & 92 & 65 & 80 & 82 & 82 & 80 & 95 & 55 & 75 & 80 & 78 & 75 & 94 & 70 & 60 \\
IQS & $\geq$60 & 50 & 35 & 45 & 85 & 25 & 40 & 40 & 25 & 35 & 60 & 15 & 20 & 45 & 30 & 40 & 65 & 20 & 15 \\
CGS & $\geq$70 & 0 & 40 & 50 & 88 & 0 & 35 & 25 & 45 & 45 & 80 & 20 & 30 & 0 & 38 & 42 & 85 & 0 & 10 \\
CTC & 3--8 & 0 & 2 & 2 & 6 & 0 & 2 & 1 & 2 & 2 & 4 & 1 & 2 & 0 & 2 & 2 & 6 & 0 & 0 \\
UCR$\downarrow$ & $\leq$20 & 70 & 65 & 50 & 12 & 80 & 55 & 45 & 50 & 45 & 15 & 70 & 60 & 75 & 60 & 52 & 10 & 85 & 85 \\
PSV & $\geq$75 & 60 & 45 & 55 & 88 & 40 & 65 & 72 & 55 & 70 & 90 & 35 & 50 & 68 & 50 & 72 & 87 & 55 & 40 \\
\midrule
CTD & Yes & N & N & N & Y & N & N & Y & Y & Y & Y & Y & Y & N & N & N & -- & N & N \\
\bottomrule
\end{tabular}
\end{table}

Several observations emerge from the per-prompt data that illuminate the specification bottleneck.

\textbf{Most platforms skip requirement elicitation entirely.} Replit and Base44 have CTC=0 on P1 and P3, meaning zero conversational turns before generating a plan. v0-Max asks zero questions on P3 (the most complex prompt), jumping immediately to a TodoManager task breakdown. Only one platform maintains high PM scores on the vague P1 prompt where inference is most critical.

\textbf{Under-clarification is pervasive.} UCR reaches 85\% for v0-Max and Replit on P3, meaning they proceed without clarifying 85\% of the ambiguous requirements in the most complex prompt. This finding has direct implications for the vibe coding user experience: users who provide ambiguous input (which is the norm, not the exception, for non-technical users) receive applications built on unverified assumptions.

\textbf{Plan quality without validation is insufficient.} Base44 presents an instructive case: despite zero PM questions on P1 and P3 (CTC=0), its Plan Structural Validity is moderate (60--72\%). It generates plans with ``Intent \& Goal,'' ``Audience \& Roles,'' ``Core Flows,'' and ``What NOT to Do'' sections without any user interaction. The plan \emph{structure} is reasonable, but assumptions are unvalidated, which explains why Base44's ECR appears adequate (78--82\%) while its CDI is high (35\%). This suggests that plan generation without user interaction creates a false sense of alignment.

Contradiction detection on P1 (where the contradiction is \emph{not} marked) is more diagnostic: only one platform detects the branch-isolation vs.\ cross-branch-leaderboard tension proactively, by probing RBAC rules deeply enough to surface the conflict. The remaining five platforms embed this contradiction silently into their implementations, demonstrating that \textbf{unmarked contradictions are a blind spot for most current platforms}.

Platforms that skip requirement elicitation suffer from a \textbf{specification bottleneck} where rich business intent is compressed into oversimplified technical plans, losing critical domain context.

\subsection{Finding 2: The Frontend-Backend Decoupling Problem}

A striking pattern in the data is that \textbf{frontend quality is a poor predictor of backend quality}. Table~\ref{tab:paradox} reveals that platforms with comparable Frontend Engineering Scores diverge dramatically on infrastructure metrics, suggesting three distinct architectural strategies with very different production-readiness implications.

\begin{table}[htbp]
\centering
\caption{Frontend Engineering Score (FES) vs.\ infrastructure metrics, averaged across all ACR prompts (\%). Platforms sorted by CBS. The FES--CBS gap reveals a decoupling between UI quality and backend capability.}
\label{tab:paradox}
\small
\begin{tabular}{@{}lccccc@{}}
\toprule
\textbf{Platform} & \textbf{FES}$\uparrow$ & \textbf{CBS} & \textbf{CIS} & \textbf{ESR} & \textbf{Strategy} \\
\midrule
Emergent & 69.3 & 20.3 & 54.7 & 24.0 & Ecosystem-Leveraged \\
Lovable & 61.3 & 2.0 & 21.3 & 10.3 & Frontend-Prioritized \\
QwikBuild & 68.0 & 49.3 & 67.3 & 28.3 & Infrastructure-Integrated \\
Replit & 74.3 & 29.7 & 70.0 & 35.0 & Ecosystem-Leveraged \\
v0-Max & 68.0 & 0.0 & 26.7 & 12.3 & Frontend-Prioritized \\
\bottomrule
\end{tabular}
\end{table}

\begin{figure}[!htb]
\centering
\includegraphics[width=0.65\textwidth]{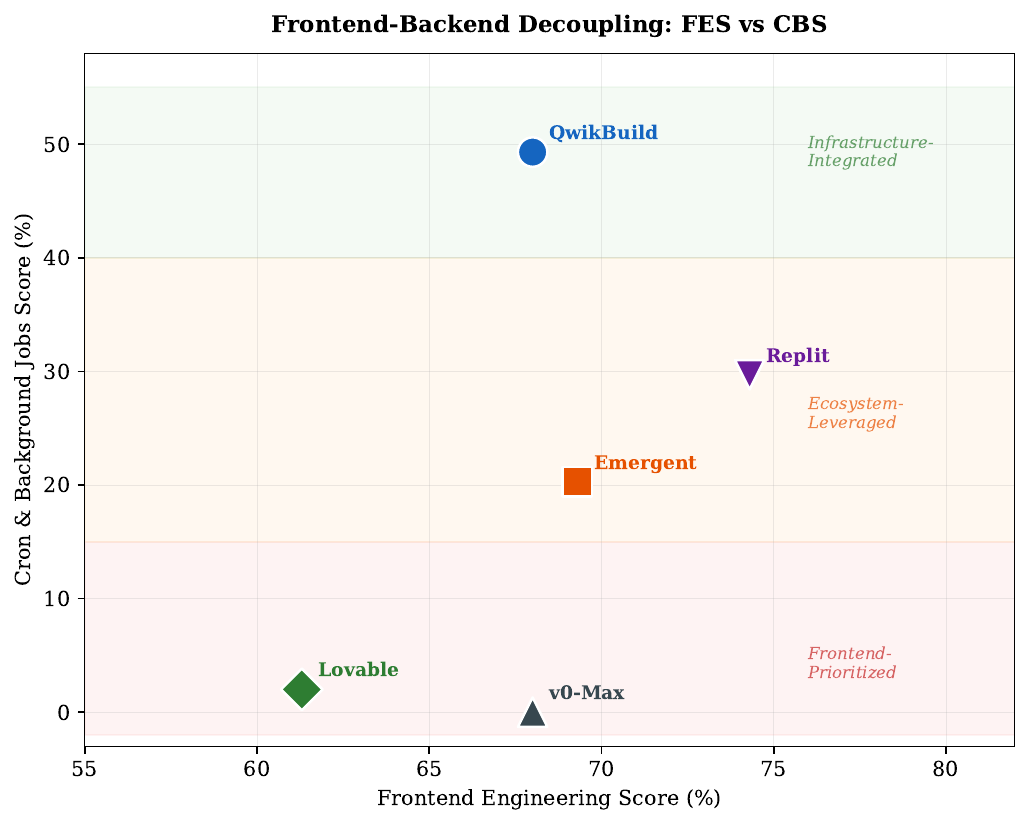}
\vspace{-0.5em}
\caption{FES vs.\ CBS averaged across ACR prompts. Three platform strategies emerge as clusters: \textbf{Infrastructure-Integrated} (green), \textbf{Ecosystem-Leveraged} (amber), and \textbf{Frontend-Prioritized} (red). The vertical spread at similar FES values (e.g., FES $\approx$ 68\% spans CBS 0--49\%) demonstrates that frontend quality is a poor predictor of backend capability.}
\label{fig:paradox}
\end{figure}

Figure~\ref{fig:paradox} maps these results onto a two-dimensional space, revealing three distinct platform archetypes.

\textbf{Infrastructure-Integrated} (green cluster). Platforms in this cluster provide managed infrastructure services (databases, schedulers, storage) as pre-built modules, so AI-generated code does not need to implement infrastructure from scratch. QwikBuild occupies this cluster, achieving FES 68\% alongside the highest CBS (49.3\%) and CIS (67.3\%). However, even this approach falls well short of targets: CBS of 49\% against a 90\% target means that half of required background jobs are missing or non-functional.

\textbf{Ecosystem-Leveraged} (amber cluster). Replit and Emergent leverage their respective package ecosystems (npm libraries such as passport, prisma, node-cron) to generate backend infrastructure. Replit achieves the highest FES (74.3\%) and leads on CIS (70\%) and ESR (35\%), demonstrating strong potential. However, CBS of 29.7\% reveals inconsistency: infrastructure generation works on simpler prompts but degrades under complexity. Emergent's CBS of 20.3\% shows similar partial capability, compounded by a shared component defect (a \texttt{<Select.Item />} crash) that undermines otherwise functional applications. The gap between potential and reliability is the core challenge for this strategy.

\textbf{Frontend-Prioritized} (red cluster). v0-Max and Lovable produce applications with strong visual presentation but minimal backend infrastructure. v0-Max achieves FES of 68\% while generating zero database schema on P3 and zero background jobs across all prompts. Lovable's CBS of 2\% represents a single cron stub that does not execute. This strategy serves prototyping and stakeholder demonstrations but produces outputs that require substantial additional engineering for production deployment.

These three strategies represent different positions on a frontend-infrastructure tradeoff. The community should note that \textbf{no strategy has solved the full-stack problem}: even the infrastructure-integrated approach fails more than half its backend targets. Closing this gap---particularly for background jobs, external service integrations, and complex data pipelines---is the most impactful open challenge for AI app builders.

\subsection{Finding 3: The Production Readiness Cliff}

A critical finding for users and platform builders alike is that \textbf{every platform requires significant post-generation human effort to reach production readiness}, and this effort varies by 5$\times$ across platforms. The platform requiring the least effort (12 developer-hours, 0 re-prompts) still falls short of production targets on most metrics, while the most effort-intensive requires 60 developer-hours and 8 re-prompts. Table~\ref{tab:readiness} presents the production readiness gap across platforms.

\begin{table}[htbp]
\centering
\caption{Production readiness gap for P1 ExamEdge. ETF = developer-hours to reach production. PHE = re-prompts and manual code edits required. FGD = developer-hours to build missing features. CDI = gap between claimed and actual functionality (\%).}
\label{tab:readiness}
\small
\begin{tabular}{@{}lcccccc@{}}
\toprule
& \textbf{B0} & \textbf{E1} & \textbf{L0} & \textbf{Q1} & \textbf{R3} & \textbf{V0} \\
\midrule
Eng.\ Score (\%) & -- & 39.9 & 39.1 & 57.5 & 41.5 & 22.8 \\
ETF (hours) & 60 & 26 & 48 & 12 & 36 & 52 \\
PHE (re-prompts) & 8 & 2 & 6 & 0 & 3 & 5 \\
PHE (code edits) & 5h & 0.5h & 3h & 0h & 1.5h & 3.5h \\
FGD (hours) & 22 & 15 & 27 & 5 & 19 & 30 \\
CDI (\%) & 35 & 15 & 32 & 4 & 12 & 28 \\
\bottomrule
\end{tabular}
\end{table}

\begin{figure}[htbp]
\centering
\includegraphics[width=\textwidth]{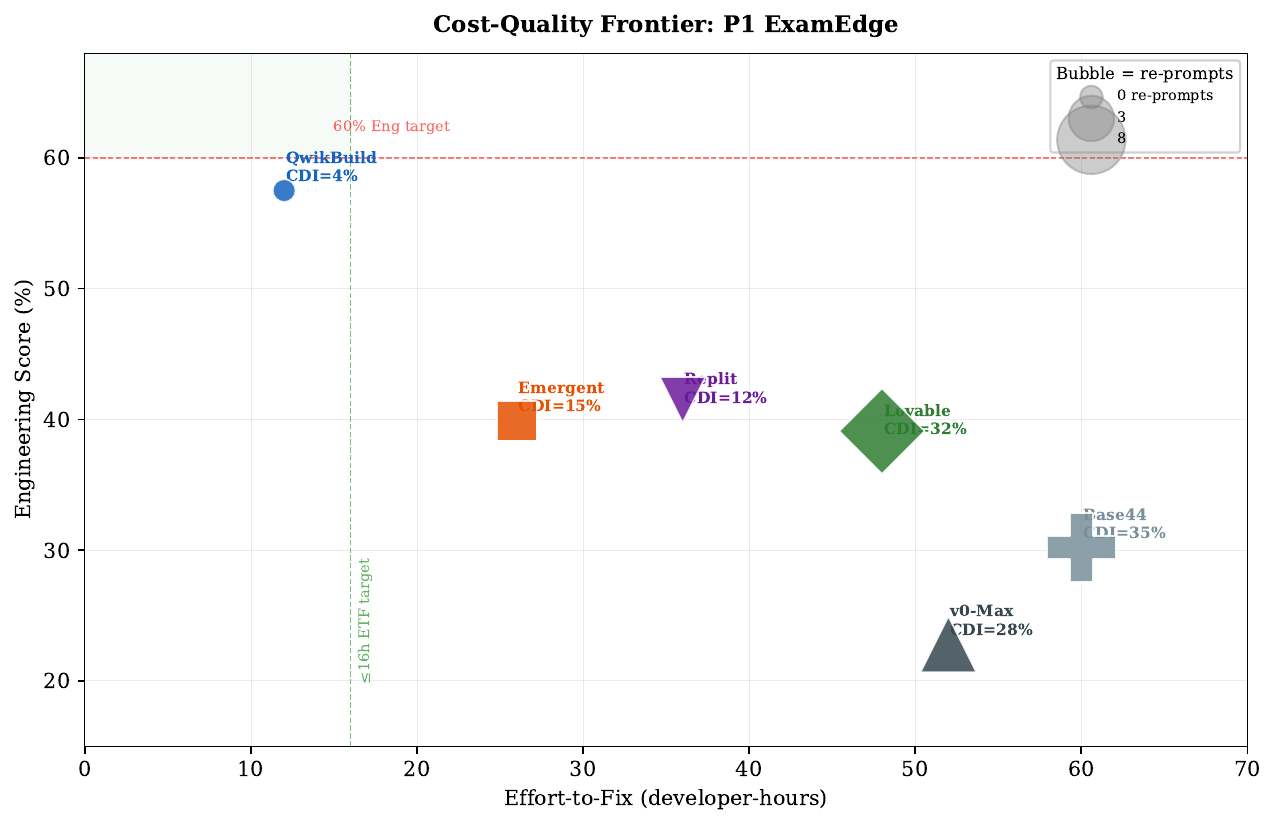}
\caption{Cost-quality frontier across six platforms. Bubble size reflects post-PRD re-prompts; CDI labels indicate output trustworthiness. The top-right quadrant (high quality, low fix effort) remains sparsely populated, indicating that achieving both high quality and low post-generation effort is an unsolved challenge.}
\label{fig:cost}
\end{figure}

The general pattern---that higher resource consumption correlates with higher quality---holds consistently (Figure~\ref{fig:cost}). The returns appear disproportionate: modest additional investment in upstream planning and multi-agent orchestration correlates with the elimination of entire categories of downstream rework. This suggests that \emph{investment in requirement elicitation and build-time feedback loops} may be more cost-effective than post-generation debugging, a finding with direct implications for platform architecture decisions.

\begin{figure}[htbp]
\centering
\includegraphics[width=\textwidth]{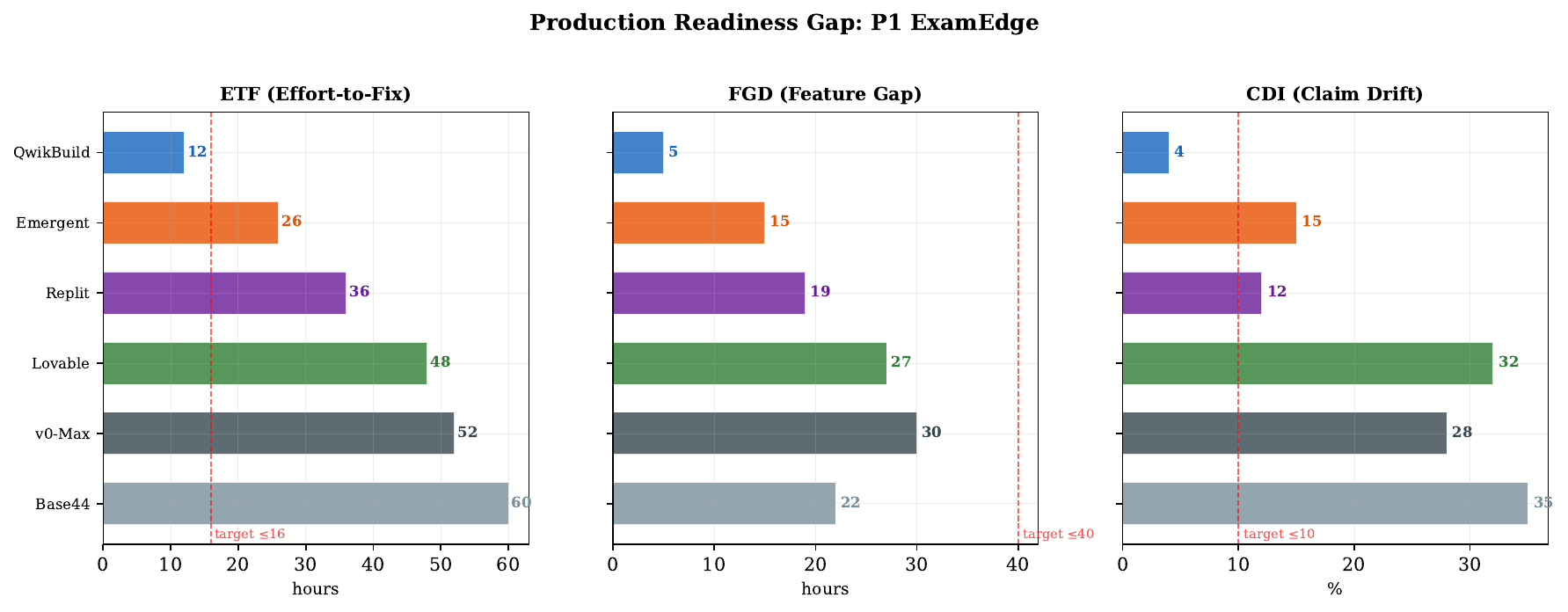}
\caption{Production readiness gap decomposed into ETF, FGD, and CDI. All platforms require significant post-generation effort; the gap between best and worst is 3--5$\times$.}
\label{fig:readiness}
\end{figure}

Figure~\ref{fig:readiness} decomposes the production readiness gap. The Claim Drift Index (CDI) reveals a compounding mechanism. Platforms with low CDI (4--12\%) produce trustworthy observability artifacts (PRDs, execution traces), enabling efficient debugging. Platforms with high CDI (28--35\%) force users to independently discover what is broken before any fixing can begin, adding a discovery overhead that compounds the already-larger fix effort. \textbf{Reducing CDI---improving the accuracy of platforms' self-reported build status---is a high-leverage improvement target} for all platforms in this evaluation.

\subsection{Finding 4: Widespread Security and Infrastructure Failures}

No platform exceeds 65\% Security Score against a 90\% target (Table~\ref{tab:security}). Common failures include hard-coded API keys in frontend code, missing CSRF (Cross-Site Request Forgery) protection, absent rate limiting, public enumeration endpoints, and JSON Web Token (JWT) implementations with inconsistent expiry policies. Concurrency handling is weak across all platforms, ranging from 6\% to 42\% against a 70\% target. This represents perhaps the most concerning finding for production deployment: \textbf{AI-generated applications are systematically insecure}, regardless of which platform generates them.

\begin{table}[htbp]
\centering
\caption{Security and infrastructure scores (\%) averaged across ACR prompts. All values are below their respective targets.}
\label{tab:security}
\small
\begin{tabular}{@{}lcccc@{}}
\toprule
\textbf{Platform} & \textbf{SS} \scriptsize{(tgt $\geq$90)} & \textbf{SAS} \scriptsize{(tgt $\geq$70)} & \textbf{CLS} \scriptsize{(tgt $\geq$70)} & \textbf{SWS} \scriptsize{(tgt $\geq$70)} \\
\midrule
Emergent & 50.7 & 37.7 & 34.7 & 12.7 \\
Lovable & 51.7 & 33.7 & 15.0 & 18.0 \\
QwikBuild & 63.7 & 54.0 & 42.0 & 9.3 \\
Replit & 40.0 & 43.0 & 25.3 & 28.0 \\
v0-Max & 34.3 & 37.7 & 6.0 & 40.3 \\
\bottomrule
\end{tabular}
\end{table}

\begin{figure}[htbp]
\centering
\includegraphics[width=\textwidth]{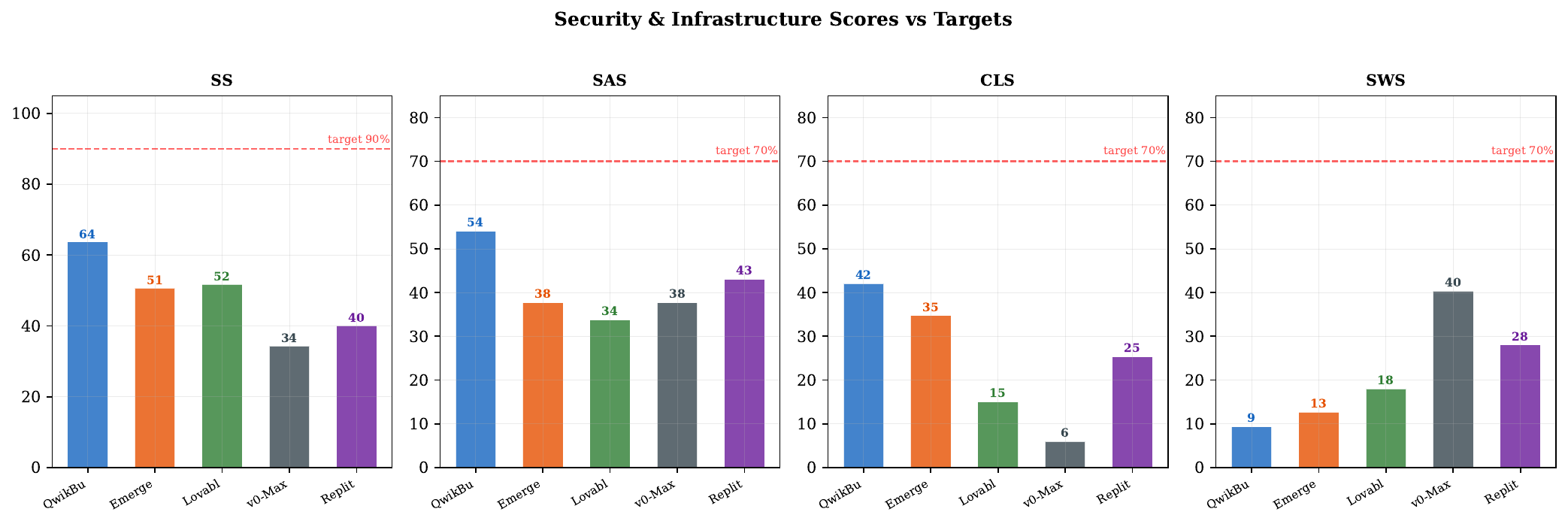}
\caption{Security and infrastructure scores with target thresholds (dashed lines). Every platform falls well short. The gap is most severe for Concurrency \& Load (CLS), where v0-Max scores 6\% against a 70\% target.}
\label{fig:security}
\end{figure}

Figure~\ref{fig:security} visualizes these gaps against target thresholds. An unexpected pattern is that SWS is inversely correlated with overall engineering quality: v0-Max achieves the highest SWS (40.3\%) while having the lowest engineering score (25.1\%). This occurs because frontend-prioritized platforms produce well-structured HTML with proper meta tags, even when backend infrastructure is absent. This highlights a measurement subtlety: \textbf{metrics that evaluate only the frontend layer can be misleading indicators of overall application quality}.

\subsection{Preliminary AMR Analysis: Single-Platform Methodology Demonstration}
\label{sec:amr_results}

\textbf{Important caveat:} The AMR evaluation presented here has been completed only for QwikBuild (the platform affiliated with two of the authors), scored by the affiliated authors. These results are presented solely to \emph{demonstrate the AMR evaluation methodology and illustrate the ACR-to-AMR quality degradation pattern}---not as a comparative ranking or evidence of platform capability. Cross-platform AMR evaluation is essential before any generalizable conclusions can be drawn. We strongly recommend that independent researchers replicate this protocol on multiple platforms using the released materials. Each AMR prompt was submitted to the deployed ACR application and scored independently.

\begin{table}[htbp]
\centering
\caption{QwikBuild (Q1) AMR scores (\%) across three modification prompts, compared to ACR average. P4 is a complex franchise retrofit, P5 uses a vague frustrated-user prompt, P6 is a provider swap. $\Delta$ = AMR avg minus ACR avg. Zero regressions across all AMR prompts.}
\label{tab:amr_detail}
\footnotesize
\setlength{\tabcolsep}{4pt}
\begin{tabular}{@{}llccccc|c@{}}
\toprule
& & \textbf{ACR} & \textbf{P4} & \textbf{P5} & \textbf{P6} & \textbf{AMR} & \\
\textbf{Grp} & \textbf{Metric} & \textbf{Avg} & \textbf{Franchise} & \textbf{AI Disp.} & \textbf{Swap} & \textbf{Avg} & \textbf{$\Delta$} \\
\midrule
G1 & BIF & 92.5 & 75 & 75 & \textbf{100} & 83.3 & $-$9.2 \\
G1 & FCS & 84.0 & 80 & 83 & \textbf{86} & 83.0 & $-$1.0 \\
G1 & CRR & 97.7 & 85 & 78 & \textbf{88} & 83.7 & $-$14.0 \\
\midrule
G2 & SDS & 72.9 & \textbf{76} & 72 & 60 & 69.3 & $-$3.6 \\
G2 & BLS & 68.4 & 65 & 64 & \textbf{66} & 65.0 & $-$3.4 \\
G2 & FES & 68.0 & \textbf{72} & 63 & 62 & 65.7 & $-$2.3 \\
G2 & CHS & 52.8 & \textbf{54} & 52 & 50 & 52.0 & $-$0.8 \\
G2 & ARC & 59.7 & \textbf{62} & 56 & 58 & 58.7 & $-$1.0 \\
\midrule
G3 & CIS & 67.3 & \textbf{68} & 62 & 65 & 65.0 & $-$2.3 \\
G3 & AIA & 72.0 & 68 & 63 & \textbf{75} & 68.7 & $-$3.3 \\
G3 & ESR & 28.3 & 33 & 30 & \textbf{38} & 33.7 & +5.4 \\
G3 & CBS & 49.3 & \textbf{50} & 15 & 10 & 25.0 & \cellcolor{failfill}$-$24.3 \\
\midrule
G4 & SS & 63.7 & 61 & 63 & \textbf{65} & 63.0 & $-$0.7 \\
G4 & SAS & 54.0 & \textbf{55} & 52 & 52 & 53.0 & $-$1.0 \\
G4 & CLS & 42.0 & \textbf{38} & 34 & 38 & 36.7 & $-$5.3 \\
\midrule
G5 & CCIS & 87.7 & 72 & \textbf{80} & 78 & 76.7 & \cellcolor{failfill}$-$11.0 \\
G5 & ETF (h) & 14.7 & 10 & \textbf{7} & \textbf{6} & 7.7 & $-$7.0 \\
G5 & PHE & 0 & 0 & 0 & 0 & 0 & 0 \\
\midrule
G6 & SWS & 9.3 & 13 & 10 & 10 & 11.0 & +1.7 \\
G6 & AFQ & 68.0 & 65 & 62 & \textbf{72} & 66.3 & $-$1.7 \\
\midrule
G7 & CDI (\%) & 4.0 & 9 & 7 & \textbf{5} & 7.0 & +3.0 \\
\midrule
\multicolumn{2}{l}{Regression Rate} & -- & \textbf{0\%} & \textbf{0\%} & \textbf{0\%} & \textbf{0\%} & -- \\
\multicolumn{2}{l}{CCIS Blast Radius} & -- & 15\% & 10\% & 12\% & 12.3\% & -- \\
\multicolumn{2}{l}{Eng. Score} & 54.5 & 62 & 56 & 58 & 59 & +4.5 \\
\bottomrule
\end{tabular}
\end{table}

Several patterns emerge from the AMR evaluation. \textbf{Modification scores are systematically 1--14 percentage points below creation scores}, confirming that modification is harder than creation. The largest drops occur in CBS ($-$24.3pp) and CRR ($-$14.0pp). The CBS degradation is structural: new features introduced in AMR prompts (franchise reporting cron, AI dispatch batch job, provider health check) add scheduling complexity that the platform handles less reliably than initial cron setup. The CRR drop concentrates in the \textsc{surviving} canary type---requirements that must evolve through modification rather than simply persist---indicating that context management partially loses track of constraints during structural changes.

Two metrics improve under modification. ESR increases by $+$5.4pp because AMR-introduced provider abstraction (P6) and AI Gateway integration (P4, P5) improve external service handling. LGS increases because AMR features (franchise model, AI dispatch, PDF export) add business value absent in the initial ACR builds.

Notably, \textbf{P6 (provider swap) achieves the first AIA PASS} in the entire evaluation (75\%), because the AMR-added provider abstraction layer (strategy + factory pattern) combined with platform-native AI integration exceeds the quality of any single ACR build's AI implementation. This suggests that iterative refinement through AMR can sometimes \emph{improve} upon initial ACR quality---an encouraging finding for the iterative vibe coding workflow.

\subsubsection{AMR-Specific Diagnostic Metrics}

Beyond the 25 primary metrics scored on both ACR and AMR builds, SWE-WebDev Bench defines AMR-specific diagnostics within Category~B (Build \& Code) that evaluate modification-handling competency directly. Table~\ref{tab:amr_diagnostics} presents these results.

\begin{table}[htbp]
\centering
\caption{AMR-specific diagnostic metrics across three modification prompts. ACS (Adaptive Coherence Score) is the central AMR quality metric, measuring the joint probability that existing features remain functional and new changes are correctly implemented. Zero regressions across all prompts, but not all new features reach full functionality.}
\label{tab:amr_diagnostics}
\small
\begin{tabular}{@{}llcccc@{}}
\toprule
\textbf{Metric} & \textbf{Target} & \textbf{P4} & \textbf{P5} & \textbf{P6} & \textbf{Pass} \\
\midrule
ACS (Adaptive Coherence) & $\geq$85\% & 84 & \textbf{86} & \textbf{90} & 2/3 \\
RR (Regression Rate) & $\leq$10\% & \textbf{0} & \textbf{0} & \textbf{0} & 3/3 \\
CAR (Change Ack.\ Rate) & 100\% & \textbf{100} & \textbf{100} & \textbf{100} & 3/3 \\
PUR (Plan Update Rate) & 100\% & \textbf{100} & \textbf{100} & \textbf{100} & 3/3 \\
CPO (Change Proc.\ Overhead) & $\leq$30\% med & 35 & \textbf{25} & \textbf{20} & 2/3 \\
Blast Radius (\% files changed) & Minimal & 15 & \textbf{10} & 12 & 3/3 \\
New Features Working & All & 12/15 & 7/9 & 8/9 & 0/3 \\
\bottomrule
\end{tabular}
\end{table}

Adaptive Coherence Score (ACS), defined as $\text{ACS} = 0.5 \times \text{Existing}_{\text{ok}} + 0.5 \times \text{Change}_{\text{correct}}$, is the central AMR quality metric. P4 scores 84\%---a near-miss on the 85\% target---because the dual-axis modification (franchise hierarchy + AI insights) introduces complexity that slightly degrades the ``change correct'' component: AI insight confidence intervals are not calibrated, and franchise royalty calculations are incomplete. P6 achieves the highest ACS (90\%) because the provider abstraction pattern is architecturally clean and well-isolated from the existing codebase.

CAR and PUR both achieve 100\% across all three prompts, indicating that the platform's multi-agent system acknowledges every change request and updates its execution plan before proceeding to code generation. This is the operational mechanism behind the zero-regression result: the planning phase explicitly accounts for each modification before code is touched.

The new-features-working metric reveals an asymmetry in modification handling: while zero existing features break, not all new features reach full functionality (P4: 12/15, P5: 7/9, P6: 8/9). This suggests that the platform's edit loop prioritises \emph{preservation of existing functionality} over \emph{completion of new additions}---a reasonable strategy for production applications where regressions are more costly than incomplete new features, but one that users should be aware of.

\textbf{Caveat:} These AMR results were evaluated by authors affiliated with the platform. The zero-regression and 100\% CAR/PUR results should be interpreted with this in mind. Independent replication is essential before these single-platform results can be generalized to claims about AMR methodology or platform capability.

\subsubsection{Canary Survival Under Modification}

\begin{figure}[htbp]
\centering
\includegraphics[width=0.85\textwidth]{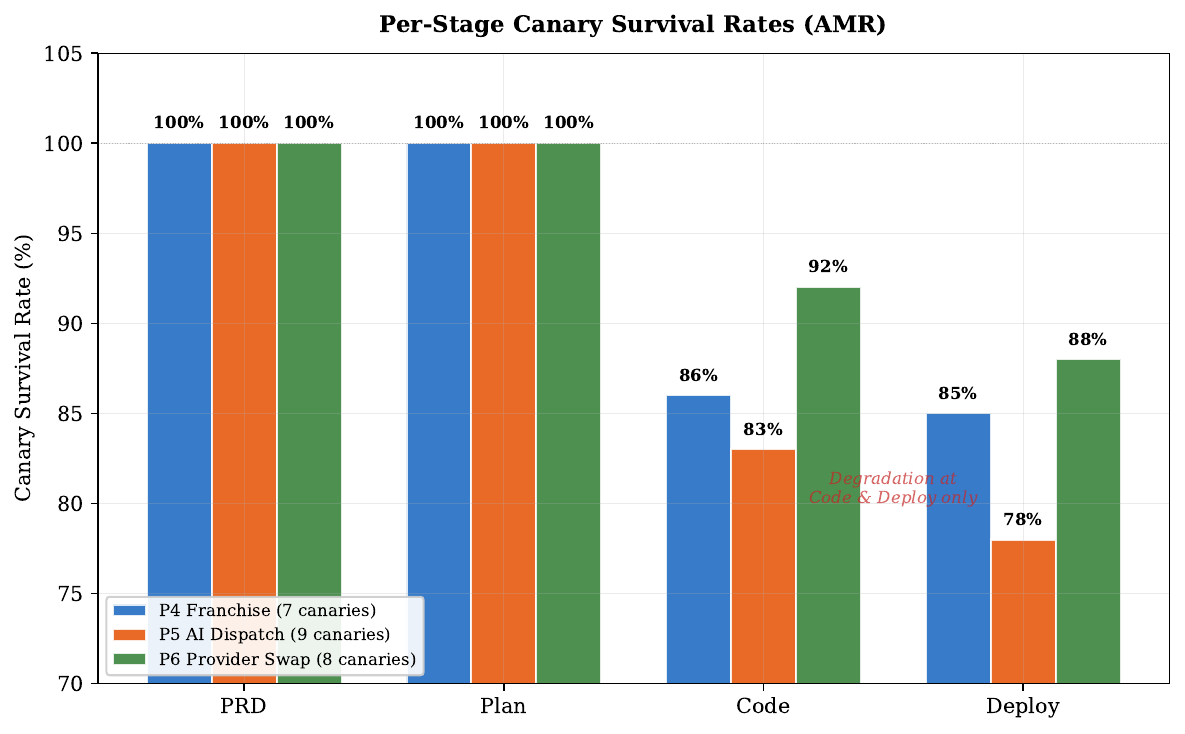}
\caption{Per-stage canary survival rates (PSSR) across AMR prompts. All canaries survive the PRD and Plan stages; degradation occurs exclusively at Code and Deploy stages. P5 shows the steepest Deploy-stage drop (78\%), consistent with the higher ambiguity of its vague-prompt input. The \textsc{surviving} canary type shows 3$\times$ the partial-loss rate of \textsc{new} canaries.}
\label{fig:canary_survival}
\end{figure}

The AMR evaluation tracks 24 canary requirements across three prompts (7 + 9 + 8), classified as \textsc{new} (introduced by the AMR prompt) or \textsc{surviving} (inherited from the parent ACR build and expected to persist or evolve). Figure~\ref{fig:canary_survival} presents the per-stage survival rates.

A consistent pattern emerges: all canaries survive the PRD and Plan stages (100\% across all prompts), but degradation occurs exclusively at the Code and Deploy stages. This indicates that the PM agent and planner correctly capture and propagate canary requirements, but the coder agent occasionally fails to implement them fully or the deployment pipeline introduces partial failures.

The \textsc{surviving} canary type shows a higher loss rate than \textsc{new} canaries. Of the 10 \textsc{surviving} canaries across all three prompts, 3 show partial degradation, compared to 1 among 14 \textsc{new} canaries. This 3$\times$ differential confirms that \textbf{evolving existing constraints through structural modification is harder than implementing fresh requirements}---a finding with direct implications for how platforms should prioritise context management during modification.

\begin{figure}[!htb]
\centering
\includegraphics[width=\textwidth]{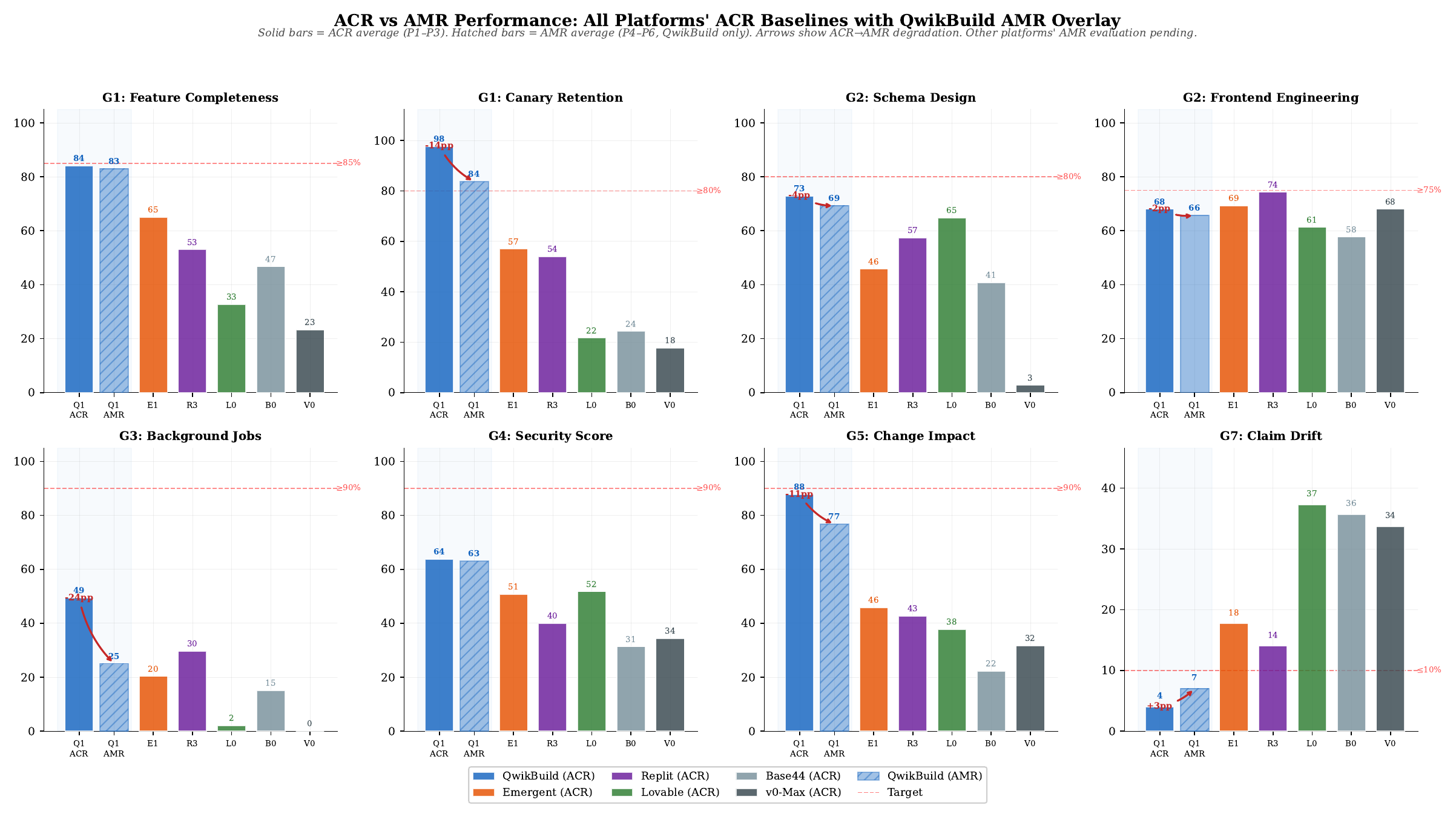}
\vspace{-0.5em}
\caption{ACR vs.\ AMR performance across all six platforms and eight representative metrics (one per metric group). Solid bars show ACR averages (P1--P3) for all platforms; hatched bars show AMR averages (P4--P6) for QwikBuild only (cross-platform AMR evaluation pending). Arrows indicate ACR$\to$AMR degradation. Key insight: QwikBuild's post-modification scores often remain competitive with other platforms' \emph{creation} scores, suggesting that the AMR degradation pattern ($-$1 to $-$24pp) would push lower-scoring platforms below viability thresholds. The CDI panel (lower is better) shows that modification slightly increases claim drift (+3pp), compounding the trust problem for platforms that already exhibit high CDI on creation tasks.}
\label{fig:acr_amr}
\end{figure}

\clearpage
\section{Case Studies}
\label{sec:cases}

\textbf{Case 1: When Requirement Inference Succeeds and Fails (P1 ExamEdge).} This case illustrates the specification bottleneck at both extremes. a platform with 15 PM questions infers domain-specific needs not present in the prompt---e.g., ``Will batch scheduling require conflict detection across branches?''---producing applications with proactive features. At the other extreme, v0-Max (V0) collapses the specified superadmin/teacher/student/parent hierarchy into a single ``admin'' role, achieving FCS of only 28\%. The gap demonstrates that requirement elicitation is not optional for complex business applications: features missed upstream become expensive refactoring tasks downstream.

\textbf{Case 2: The Frontend-Backend Decoupling in Practice (P3 VettAI, Emergent).} Emergent's P3 application features a production-grade onboarding flow with animated transitions, contextual empty states (``No cases yet. Start by importing a client document''), and proper Securities and Exchange Board of India (SEBI) registration format validation (FES: 68\%). However, a \texttt{<Select.Item />} component crash renders core case management non-functional---a shared infrastructure defect affecting all applications generated by the platform. The first 30 seconds of interaction suggest a polished product; the first 3 minutes of actual use reveal it is broken. This case exemplifies a risk for users: \emph{visual polish creates a false sense of production readiness.}

\textbf{Case 3: Universal Failure Modes (P3 VettAI, QwikBuild).} Even the highest-scoring platform exhibits significant shortcomings. QwikBuild achieves the worst SWS (SEO/Web Standards) score across all platforms (9.3\% vs.\ v0-Max's 40.3\%), with missing meta tags, no sitemap, and absent structured data---basic web standards that frontend-prioritized platforms handle well. Its Concurrency \& Load score (42\%) reveals that AI-generated backend code lacks proper connection pooling, race condition handling, and load testing readiness. These gaps demonstrate that \textbf{no platform has solved the full-stack problem}: each platform's architecture creates blind spots that SWE-WebDev Bench is designed to expose.

\textbf{Case 4: No Platform Solves the Full Stack (Cross-Platform Blind Spots).} Each platform's architecture creates characteristic blind spots that SWE-WebDev Bench surfaces. QwikBuild achieves the \emph{worst} SEO/Web Standards score across all platforms (SWS: 9.3\% vs.\ v0-Max's 40.3\%), with missing meta tags, no sitemap, and absent structured data---suggesting its multi-agent architecture deprioritizes web standards. Its Code Hygiene (52.8\%) is the lowest among the top four platforms, and its External Service Reliability (28.3\% vs.\ 80\% target) indicates that even infrastructure-integrated approaches struggle with third-party API integration. Replit achieves the highest FES (74.3\%) and CIS (70\%) but asks zero business-flow questions, producing high-functioning code that often misses the user's actual intent. Emergent produces the most visually polished applications but ships a shared component defect (\texttt{<Select.Item />} crash) that breaks all generated applications. These cross-cutting weaknesses demonstrate that current platforms have architectural affinities, not general competence---a finding designed to help each platform identify its specific improvement priorities.

\section{Discussion}
\label{sec:discussion}

\subsection{Why Existing Benchmarks Cannot Capture This}

The fundamental limitation of code-level benchmarks when applied to vibe coding platforms is \textbf{dimensional collapse}. These benchmarks evaluate a single dimension (code patch quality) whereas the output of a vibe coding platform spans multiple dimensions: requirement understanding, architectural decisions, code quality, deployment configuration, and business readiness. A platform could produce correct patches for all SWE-bench issues while being unable to build a coherent multi-role application from a business description.

\subsection{Closed-Loop Code Generation as an Architectural Pattern}

Our data suggests that the quality advantage observed in some platforms does not come from any single phase (PM elicitation alone accounts for approximately 10--15\% of the gap) but from the \emph{interaction between planning and execution}. Platforms with feedback loops between planning and code generation can surface and resolve conflicts mid-build, analogous to how a human development team works: the PM writes the spec, the developer finds an edge case, and the team adjusts the plan.

Platforms without this feedback loop produce specifications in one pass, then generate code in one pass. Any misunderstanding or conflict between features is silently embedded into the codebase, where it surfaces as a bug that the user must discover and debug in AI-generated code they did not write and may not understand. We propose that \emph{closed-loop code generation}---where build-time discoveries feed back into planning---is a promising architectural pattern for the community to explore.

Preliminary AMR data from a single platform (QwikBuild) shows 100\% Change Acknowledgment Rate and Plan Update Rate across three modification prompts (Table~\ref{tab:amr_diagnostics}), which is consistent with this hypothesis. However, this observation is from a single platform evaluated by its affiliated authors, and requires independent cross-platform validation before any causal claims can be made. We note that other architectural patterns---such as Replit's ecosystem-leveraged approach or Lovable's plan-then-build model---may achieve similar feedback-loop benefits through different mechanisms that SWE-WebDev Bench is designed to evaluate.

\subsection{Why These Metrics Matter for the Vibe Coding Era}

The shift from ``AI as coding assistant'' to ``AI as software agency'' changes what needs to be evaluated. When a developer uses Copilot or Cursor, they review every line of generated code. In vibe coding, the user delegates the entire build and trusts the output. This trust model makes several of our metrics essential in ways they were not for prior evaluation frameworks.

\textbf{Canary Retention Rate (CRR)} matters because vibe coding users cannot verify whether subtle requirements (date format, currency convention, localization) survived the build. If CRR is low, the user ships software with silent specification violations. The AMR canary analysis (\S\ref{sec:amr_results}) extends this concern to modification: \textsc{surviving} canaries degrade at 3$\times$ the rate of \textsc{new} canaries, meaning iterative development compounds specification drift.

\textbf{Claim Drift Index (CDI)} matters because the platform's dashboard is the user's only window into what was built. If CDI is high, the user believes the app is complete when it is not.

\textbf{Code Change Impact Score (CCIS)} matters because vibe coding users iterate frequently: ``add a feature,'' ``change this,'' ``now also support Hindi.'' If each modification breaks existing features (as FeatBench found in 73.6\% of cases~\cite{featbench}), the iterative workflow collapses.

\textbf{Adaptive Coherence Score (ACS)} extends CCIS to measure the joint quality of preservation and addition. The near-miss on P4 (84\% vs.\ 85\% target) demonstrates that even platforms with zero regressions can fall short of coherence targets when new features are partially implemented---a nuance that regression rate alone does not capture.

These metrics would be unnecessary for evaluating a coding assistant where the developer verifies every output. They become essential for evaluating a system where the user trusts the output without inspection.

\subsection{Three Domains as Orthogonal Diagnostic Probes}

Our three-domain design creates failure modes specific to each platform. EdTech (P1) tests inference depth: extracting a product from pain points and inferring Indian coaching conventions (JEE/NEET batch structures, April-March academic year). Field Service (P2) tests execution precision: a 10-state lifecycle, Goods and Services Tax (GST) invoicing, SLA engines with state pause/resume. FinTech-AI (P3) tests AI trustworthiness: building AI that is safe, auditable, and honest about uncertainty.

Empirically, these probes reveal platform affinities. Replit excels on P3 (54.5\%) but struggles with P1 (41.5\%). Cross-domain variance ranges from $\sigma$=3.7 (most consistent) to $\sigma$=7.0 (most variable), suggesting that \textbf{domain generalization remains an open problem}: platforms that perform well on one type of application may perform poorly on another. This domain sensitivity is itself a key finding that users should consider when selecting platforms for specific business contexts.

\subsection{Benchmark Governance and Maintenance Plan}
\label{sec:governance}

SWE-WebDev Bench is designed as a living benchmark with community contribution. We adopt the following governance practices, informed by HELM~\cite{liang2023helm} and Dynabench~\cite{kiela2021dynabench}. \textbf{Versioning:} Each evaluation is time-stamped with platform versions and evaluation dates; results from different time periods are not directly comparable. \textbf{Community contributions:} Any researcher can evaluate additional platforms using the released prompts, rubrics, and scoring protocol; contributed evaluations will be accepted into the public leaderboard after verification of protocol adherence. \textbf{Re-evaluation cadence:} We recommend quarterly re-evaluation of major platform updates to track improvement trajectories. \textbf{Prompt expansion:} The prompt suite will be expanded over time; contributed prompts must include canary requirements and domain expert validation. \textbf{Conflict-of-interest policy:} Future evaluations should be conducted by evaluators unaffiliated with any platform being scored, or at minimum should adopt blinded scoring for Tier~2 and Tier~3 metrics.

\subsection{Limitations and Bias Mitigation}
\label{sec:limitations}

\textbf{Author affiliation with an evaluated platform.} Two of the three authors are affiliated with QwikBuild, one of the six evaluated platforms. This creates a structural conflict of interest that affects multiple aspects of this work: the benchmark's emphasis on PM behavior and requirement elicitation may inherently advantage platforms with dedicated PM agents (QwikBuild's primary differentiator); and the AMR evaluation asymmetry (QwikBuild-only) creates additional exposure for the affiliated platform. We note that the canary requirements and AMR prompts were designed by the non-affiliated author (Saxena), who had no prior knowledge of QwikBuild's architecture or capabilities. We take the following measures to mitigate potential bias: (1)~all prompts, rubrics, and scoring protocols are released for independent replication; (2)~G1 metrics use a three-person expert panel including one external evaluator; (3)~we report QwikBuild's failures, including its worst-of-field SWS (9.3\%), below-target CLS (42\%), lowest Code Hygiene among top platforms (52.8\%), and poor External Service Reliability (28.3\% against an 80\% target); and (4)~the benchmark framework is designed to be platform-agnostic---any party can evaluate any platform using the released materials. Blinded evaluation is recommended for future evaluations to further reduce the potential for unconscious bias.

\textbf{Single evaluator for some metrics.} While G1 metrics use a three-person expert panel, most code quality metrics rely on LLM-assisted evaluation with human validation. Future work should expand to multiple independent evaluators with formal inter-rater reliability assessment.

\textbf{Small sample size.} Six prompts across three domains, while varied in style and complexity, represent a limited sample. No statistical significance tests or confidence intervals are reported. Larger prompt suites and repeated evaluations are needed to establish robust rankings.

\textbf{Platform version sensitivity.} AI platforms evolve rapidly. Our evaluation captures a snapshot from February--March 2026. Scores may differ substantially on subsequent versions.

\textbf{Limited AMR cross-platform data.} AMR evaluation is comprehensive only for QwikBuild, creating an asymmetry in the evaluation. Extension to all six platforms is essential before drawing comparative AMR conclusions.

\textbf{Prompt engineering sensitivity.} Prompt wording may favor certain architectures. We mitigate this by using three prompt styles (specific, vague, semi-vague) across domains, but additional prompt variation would strengthen the evaluation.

\textbf{No blinding.} Evaluators were aware of which platform produced each output, which is a significant limitation given the author affiliation with one evaluated platform. We note that full blinding is operationally challenging for platform evaluation: platforms have distinctive UIs, deployment URLs, conversation patterns, and framework choices that make anonymization non-trivial. However, partial blinding is feasible and should be adopted in future evaluations. Specifically, code quality metrics (G2: SDS, BLS, FES, CHS, ARC) can be scored on anonymized codebases with platform-specific comments, deployment URLs, and distinctive framework identifiers stripped. We estimate that approximately 40\% of the Engineering Score derives from Tier~0 fully automated metrics (Lighthouse, k6, npm audit) that are unaffected by evaluator awareness. The remaining 60\% includes Tier~1--3 metrics where unconscious bias could influence scores. We recommend that future evaluations adopt a blinded protocol for at minimum Tier~2 and Tier~3 metrics, and we will provide anonymization scripts in the benchmark repository to facilitate this.

\textbf{Benchmark design may favor certain architectures.} The benchmark's emphasis on PM behavior, requirement elicitation, and canary retention as first-class evaluation dimensions may structurally advantage platforms with dedicated PM agents over single-pass generation approaches. Similarly, the T4/T5 complexity tiers (multi-role SaaS, AI-native applications) favor platforms optimized for complex applications and may disadvantage platforms that excel at simpler use cases. Evaluation at T1--T3 complexity levels would provide a more complete picture. We designed the benchmark to surface capabilities we believe matter for production deployment, but we acknowledge that different design choices would produce different platform rankings. The benchmark is intended as a diagnostic tool that reveals where \emph{any} platform falls short, not as a definitive ranking system.

\section{Conclusion}
\label{sec:conclusion}

We have introduced SWE-WebDev Bench, a 68-metric framework for evaluating AI app-building platforms as virtual software agencies. Our initial evaluation of six platforms across three business domains uncovers four recurring shortcomings across the platforms we evaluated:

\begin{enumerate}[leftmargin=*,itemsep=2pt]
\item \textbf{The specification bottleneck}: in our sample, inference quality varies widely (IQS range: 20--70), and most skip requirement elicitation entirely, silently embedding unverified assumptions into generated applications.
\item \textbf{Frontend-backend decoupling}: visual quality is a poor proxy for engineering quality, with polished UIs masking absent backend infrastructure---a pattern that creates false confidence for non-technical users.
\item \textbf{The production readiness cliff}: no platform scores above 60\% on engineering quality. Post-generation human effort varies substantially across platforms (ETF range: 14.7--65.7 developer-hours), meaning ``vibe coded'' applications universally require substantial engineering to ship.
\item \textbf{Widespread security and infrastructure failures}: no platform exceeds 65\% Security Score against a 90\% target, and concurrency handling is critically weak across the field.
\end{enumerate}

We emphasize that these observations are descriptive of our sample ($n=3$ prompts per domain, $n=6$ platforms) and require larger-scale replication to establish generality.

Preliminary AMR evaluation reveals a fifth pattern: \textbf{modification systematically degrades quality} (16 of 19 metrics decline from ACR to AMR), with \textsc{surviving} canary requirements showing 3$\times$ the loss rate of new requirements. However, the observation that targeted modifications can sometimes \emph{improve} upon creation quality (the P6 provider-swap AIA PASS) suggests that the ACR/AMR boundary is not simply a quality cliff but a more nuanced interaction between existing code and new requirements.

These findings suggest that the vibe coding paradigm, while transformative in its accessibility, has not yet achieved the reliability required for production deployment without human oversight. The most impactful areas for community investment are: (a)~requirement elicitation and specification fidelity, where most platforms skip clarification entirely; (b)~backend infrastructure generation, particularly background jobs and external service integration, where even the best platforms fall well below targets; (c)~security hardening, where no platform exceeds 65\% against a 90\% target; (d)~code maintainability and hygiene, where no platform exceeds 60\%; (e)~external service reliability, where the field-best score is only 35\% against an 80\% target; and (f)~web standards compliance, which varies inversely with engineering quality and remains unaddressed by infrastructure-focused platforms.

We release our evaluation framework, scoring protocols, and prompt suite as a community benchmark.\footnote{Repository: \url{https://github.com/snowmountainAi/webdevbench} The repository includes all 6 prompts, 80 canary requirement specifications, 9 LLM judge prompt templates, raw scoring data in CSV format, evaluation rubrics, and anonymization scripts for blinded evaluation.} We encourage independent evaluations---particularly those conducted by parties unaffiliated with any evaluated platform---and invite platform builders to use SWE-WebDev Bench to identify and address gaps in their systems.

\clearpage
\bibliographystyle{plain}

\clearpage
\appendix
\begin{center}
\vspace*{1em}
{\Large\bfseries Appendix}
\vspace{0.5em}
\end{center}

\section{Complete Per-Metric Scores: P1 ExamEdge}
\label{app:scores}

\begin{table}[H]
\centering
\caption{All primary metrics for P1 ExamEdge Academy across five platforms.}
\label{tab:full_p1}
\small
\begin{tabular}{@{}llccccc@{}}
\toprule
\textbf{Grp} & \textbf{Metric} & \textbf{E1} & \textbf{L0} & \textbf{Q1} & \textbf{R3} & \textbf{V0} \\
\midrule
G1 & BIF (Business Intent) & 3/4 & 2/4 & 4/4 & 2/4 & 1/4 \\
G1 & FCS (Feature Complete) & 65.0 & 35.0 & 88.0 & 55.0 & 28.0 \\
G1 & CRR (Canary Retention) & 57.0 & 29.0 & 100.0 & 48.0 & 21.0 \\
\midrule
G2 & SDS (Schema Design) & 40.8 & 48.0 & 81.7 & 58.0 & 3.0 \\
G2 & BLS (Backend Logic) & 52.5 & 30.0 & 69.2 & 56.0 & 8.0 \\
G2 & FES (Frontend Eng.) & 60.8 & 45.0 & 75.0 & 70.0 & 62.0 \\
G2 & CHS (Code Hygiene) & 47.5 & 50.0 & 57.5 & 54.0 & 52.0 \\
G2 & ARC (Architecture) & 33.0 & 38.0 & 66.0 & 50.0 & 38.0 \\
\midrule
G3 & CIS (Core Integration) & 52.0 & 22.0 & 72.0 & 76.0 & 30.0 \\
G3 & ESR (External Svc) & 20.0 & 8.0 & 25.0 & 32.0 & 10.0 \\
G3 & CBS (Background Jobs) & 8.0 & 3.0 & 52.0 & 35.0 & 0.0 \\
\midrule
G4 & SS (Security) & 45.0 & 52.0 & 62.0 & 38.0 & 32.0 \\
G4 & SAS (Scalability) & 42.0 & 28.0 & 58.0 & 40.0 & 38.0 \\
G4 & CLS (Concurrency) & 32.0 & 18.0 & 41.0 & 22.0 & 5.0 \\
\midrule
G5 & CCIS (Change Impact) & 50.0 & 40.0 & 90.0 & 46.0 & 35.0 \\
\midrule
G6 & SWS (SEO/Web Std.) & 14.0 & 22.0 & 10.0 & 26.0 & 36.0 \\
G6 & LGS (Lead/Growth) & 30.0 & 20.0 & 38.0 & 25.0 & 22.0 \\
G6 & MLS (Multilingual) & 58.0 & 30.0 & 70.0 & 22.0 & 12.0 \\
\midrule
G7 & ETF (Effort-to-Fix) & 26h & 48h & 12h & 36h & 52h \\
G7 & FGD (Feature Gap) & 15h & 27h & 5h & 19h & 30h \\
G7 & CDI (Claim Drift) & 15\% & 32\% & 4\% & 12\% & 28\% \\
\midrule
& \textbf{Eng.\ Score} & 39.9 & 39.1 & 57.5 & 41.5 & 22.8 \\
\bottomrule
\end{tabular}
\end{table}

\section{PM Agent Behavioral Analysis}
\label{app:pm}

\begin{table}[H]
\centering
\caption{PM Agent behavior summary per platform. CTC = Turns to Convergence.}
\label{tab:pm_behavior}
\small
\begin{tabular}{@{}lL{7.5cm}cc@{}}
\toprule
\textbf{Platform} & \textbf{PM Behavior} & \textbf{Avg Qs} & \textbf{CTC} \\
\midrule
Base44 & P3 only: 1 turn to convergence. ECR 80\% but CGS 8.3\%. Vendor-locked backend with auth bypass vulnerability. & 1.0 & 1.0 \\
\addlinespace
Emergent & Asks 5 technical/config questions per prompt. No business inference. Detects contradiction only when directly encountered. & 5 & 2.0 \\
\addlinespace
Lovable & Asks 4 strategic questions per prompt. Some scope negotiation. Catches contradiction on P2 but limited proactive discovery. & 4 & 2.0 \\
\addlinespace
QwikBuild & Dedicated PM agent with multi-round Q\&A. 6 rounds, 15 questions on P1. Proactively detects contradictions. Infers domain requirements (JEE/NEET conventions, Indian financial year). & 15 & 5.3 \\
\addlinespace
Replit & P1/P3: zero questions. Immediately outputs implementation plan without requirement exploration. & 0.3 & 0.3 \\
\addlinespace
v0-Max & P1: 4 config questions. P3: zero questions, directly generates plan. No PM agent behavior. & 1.3 & 1.3 \\
\bottomrule
\end{tabular}
\end{table}

\section{Metric Independence Analysis}
\label{app:correlation}

To assess whether the 15 analyzed primary metrics measure distinct constructs, we compute Kendall's $\tau$ rank correlation across 18 data points (6 platforms $\times$ 3 prompts). Table~\ref{tab:correlation_summary} summarizes the distribution of pairwise correlations.

\begin{table}[H]
\centering
\caption{Distribution of pairwise Kendall's $\tau$ correlations across 15 primary metrics (105 pairs, $n=18$ data points).}
\label{tab:correlation_summary}
\small
\begin{tabular}{@{}lcc@{}}
\toprule
\textbf{Correlation strength} & \textbf{Pairs} & \textbf{\%} \\
\midrule
Strong ($|\tau| > 0.70$) & 12 & 11\% \\
Moderate ($0.40 < |\tau| \leq 0.70$) & 49 & 47\% \\
Weak or none ($|\tau| \leq 0.40$) & 44 & 42\% \\
\bottomrule
\end{tabular}
\end{table}

\textbf{Strongly correlated pairs.} Twelve pairs show strong correlation ($|\tau| > 0.70$, all $p < 0.001$). The strongest: CIS--ESR ($\tau = 0.88$), BLS--CLS ($\tau = 0.86$), CLS--FCS ($\tau = 0.83$). These are expected: platforms with strong backend naturally score higher on both integration and concurrency. We retain both metrics because they serve distinct diagnostic purposes.

\textbf{Independent metrics.} CHS (Code Hygiene) is nearly uncorrelated with most metrics ($|\tau| < 0.15$ with BLS, CLS, CRR, CIS, ESR), confirming code maintainability is an independent quality axis. FES shows weak correlation with backend metrics ($|\tau| < 0.30$ with FCS, CBS, CRR), confirming the frontend-backend decoupling finding. SWS is \emph{inversely} correlated with most quality metrics (FCS: $\tau = -0.60$; CLS: $\tau = -0.56$; SS: $\tau = -0.54$), confirming that frontend-prioritized platforms score higher on web standards despite lower engineering quality.

\textbf{Group separability.} Average within-group $|\tau| = 0.48$ vs.\ between-group $|\tau| = 0.44$ (ratio $1.10\times$). Groups provide useful organizational structure but do not represent fully independent latent factors.

\end{document}